\documentclass[journal]{IEEEtran}
\IEEEoverridecommandlockouts

\usepackage{cite}
\usepackage{amsmath,amssymb,amsfonts}
\usepackage{amsthm}
\usepackage{algorithm}
\usepackage{algpseudocode}
\usepackage{graphicx}
\usepackage{booktabs}
\usepackage{multirow}
\usepackage{array}
\usepackage{xcolor}
\usepackage{xurl}
\usepackage[hidelinks]{hyperref}
\usepackage{caption}

\newcommand{\figorbox}[3]{%
  \IfFileExists{#1}{\includegraphics[width=#2]{#1}}%
  {\fbox{\begin{minipage}[c][#3][c]{#2}\centering\footnotesize
   [figure placeholder]\end{minipage}}}}

\usepackage{etoolbox}

\newcommand{\EHTSDisplaySpacing}{%
  \setlength{\abovedisplayskip}{5pt plus 1pt minus 1pt}%
  \setlength{\belowdisplayskip}{5pt plus 1pt minus 1pt}%
  \setlength{\abovedisplayshortskip}{0pt plus 1pt}%
  \setlength{\belowdisplayshortskip}{4pt plus 1pt minus 1pt}%
}

\AtBeginDocument{%
  \apptocmd{\normalsize}{\EHTSDisplaySpacing}{}{}%
  \apptocmd{\small}{\EHTSDisplaySpacing}{}{}%
  \apptocmd{\footnotesize}{\EHTSDisplaySpacing}{}{}%
  \EHTSDisplaySpacing
}

\begin{document}
\bstctlcite{EHTS:BSTcontrol}

\title{Dual-Agent Reinforcement Learning for Vehicle Scheduling and Multi-Energy Dispatch in Electric–Hydrogen-Transport Systems}

\author{Fulong Yao\textsuperscript{*,\textdagger},
Yiming Xu\textsuperscript{\textdagger},
Maurizio Albano,~\IEEEmembership{Member,~IEEE},
Liana Cipcigan,~\IEEEmembership{Member,~IEEE},
Naeima Hamed, Bo Zou, and
Omer Rana,~\IEEEmembership{Member,~IEEE}%
\thanks{This work was supported in part by the CLEETS Global Centre under NSF Award No.\ 2330565 and UKRI Award No.\ EP/Y026233/1. (\textsuperscript{*}Corresponding author;
\textsuperscript{\textdagger}equal contribution.) F. Yao, N. Hamed, and O. Rana are with the School of Computational
and Mathematical Sciences, and Y. Xu, M. Albano, and L. Cipcigan
are with the School of Engineering, Cardiff University, Cardiff, U.K.
(e-mail: \{yaof, HamedNH, RanaOF, XuY130, AlbanoM,
CipciganLM\}@cardiff.ac.uk). B. Zou is with the Department of Civil,
Materials, and Environmental Engineering, University of Illinois
Chicago, Chicago, IL, USA (e-mail: bzou@uic.edu).}
}

% \author{Fulong Yao\textsuperscript{*,\textdagger}, Yiming Xu\textsuperscript{\textdagger}, Maurizio Albano,~\IEEEmembership{Member,~IEEE}, Liana Cipcigan,~\IEEEmembership{Member,~IEEE},
% \\ Naeima Hamed, Bo Zou, and Omer Rana,~\IEEEmembership{Member,~IEEE}%
% \thanks{This work was supported in part by the CLEETS Global Centre under NSF Award No.\ 2330565 and UKRI Award No.\ EP/Y026233/1.
% (\textsuperscript{*}Corresponding author: Fulong Yao.
% \textsuperscript{\textdagger}These authors contributed equally). Fulong Yao, Naeima Hamed, and Omer Rana are with the School of
% Computational and Mathematical Sciences, Cardiff University, Cardiff,
% U.K. (e-mail: \{yaof, HamedNH, RanaOF\}@cardiff.ac.uk). Yiming Xu, Maurizio Albano, and Liana Cipcigan are with the
% School of Engineering, Cardiff University, Cardiff, U.K.
% (e-mail: \{XuY130, AlbanoM, CipciganLM\}@cardiff.ac.uk). Bo Zou is with the Department of Civil, Materials, and Environmental Engineering, University of Illinois Chicago, USA. (e-mail: bzou@uic.edu).}
% }

\maketitle

\begin{abstract}
Electric–hydrogen–transport systems (EHTSs) require coordinated scheduling of electric and hydrogen vehicles together with multi-energy dispatch. These operations form a sequential decision process: upstream vehicle scheduling (VS) determines service allocation and order, thereby endogenously shaping the electricity and hydrogen demands faced by downstream energy dispatch (ED). 
% Existing hierarchical approaches commonly rely on exogenous transport demand or fixed scheduling rules, limiting joint adaptability across the two layers. 
This paper develops a dual-agent reinforcement learning framework tailored to the heterogeneous decision structures. The vehicle-scheduling agent combines proximal policy optimisation (PPO) with a Plackett–Luce policy to learn vehicle-priority permutations, which are then mapped to feasible charger and refueller assignments through a greedy allocation. The energy-dispatch agent employs twin delayed deep deterministic policy gradient (TD3) for continuous storage and energy-flow control. The directional coupling between the two decision layers is analysed, revealing that updates to the upstream scheduling policy can shift the demand distribution experienced by the downstream dispatcher. A three-stage training mechanism is therefore introduced, combining layer-wise pretraining with concurrent fine-tuning to mitigate the effects of scheduling-induced distribution shifts on downstream ED learning. Year-long case studies show that the proposed framework achieves the lowest vehicle-service delay and energy cost among the 11 evaluated methods, demonstrating effective coordination between vehicle scheduling and multi-energy dispatch.

\end{abstract}


\raggedbottom
\setlength{\textfloatsep}{6pt}
\setlength{\dbltextfloatsep}{6pt}
\setlength{\intextsep}{6pt}
\setlength{\floatsep}{4pt}
\setlength{\dblfloatsep}{4pt}
\setlength{\parskip}{0pt}

%================================================
\section{Introduction} \label{sec:1}

Multi-energy refuelling hubs that serve both battery electric vehicles (EVs) and hydrogen vehicles (HVs) from a single site are moving from concept to deployment, driven by the complementary roles of EVs and HVs in light- and heavy-duty transport, respectively \cite{aegisenergy2025hubs, cpcatapult2025hgv,iea2025ev}. The simultaneous development of battery-electric and hydrogen mobility is also motivating infrastructure that can provide charging and hydrogen refuelling services within a shared site rather than through fully separated facilities. Recent research has consequently considered hybrid or electric--hydrogen refuelling systems that integrate transport demand with electricity and hydrogen supply infrastructures \cite{chen2025superconducting,wan2025coordinated,chen2024risk,zheng2025safe}. Such an electric--hydrogen--transport system (EHTS) couples a transport service process with a local energy system comprising photovoltaic (PV) generation, batteries, electrolysers, hydrogen storage tanks, grid connections, and hydrogen pipelines. This integration creates opportunities for shared renewable utilisation and temporal energy shifting, but also makes system operation dependent on the interaction between vehicle service and multi-energy supply.

Operation of an EHTS involves two sequential decisions per control interval: vehicle scheduling (VS) and energy dispatch (ED). The former determines vehicle-service priorities and their feasible allocation to available chargers and refuellers, thereby shaping service timing, delay, and facility utilisation. The latter determines how the resulting electricity and hydrogen demands are met by storage, renewables, and external supply, which in turn affects energy cost. The ED decision therefore depends on the VS decision, as these demands are endogenously created by VS rather than drawn from an external demand profile.

Existing studies generally emphasise one side of this interaction. ED side research has developed optimisation and learning-based strategies for the operation of electricity--hydrogen systems under uncertain renewable generation, energy prices, and demand \cite{jadidbonab2026drl, huy2024robust, yao2025unified, zhang2025hydrogen, yuan2024distributed}. However, the demand faced by these energy-management models is typically predefined, forecast, or otherwise specified independently of real-time vehicle scheduling. Conversely, VS-side research has investigated EV charging allocation, queue-aware service,
refuelling scheduling, and vehicle--station interaction \cite{zhao2024two, tsaousoglou2023fair, brown2022analysis, abdelghany2025optimizing}, but commonly optimises service or mobility objectives without explicitly propagating the resulting vehicle-level decisions into a downstream electricity--hydrogen dispatch problem.

A recent hierarchical framework \cite{yao2026hierarchical} takes a step towards addressing this scheduling-to-dispatch gap by propagating scheduler-inferred demand into a deep reinforcement learning (DRL)-based dispatcher. Its VS layer employs a solver-free greedy heuristic (SFGH) to determine vehicle scheduling and infer the corresponding energy demand, while its ED layer uses twin delayed deep deterministic policy gradient (TD3) to manage battery, hydrogen-tank, and PV operation. 
The results showed that SFGH outperformed representative heuristic schedulers, and TD3 achieved the lowest energy cost among the evaluated DRL dispatchers under the same SFGH scheduling policy. This establishes strong baselines for both layers and demonstrates the benefit of explicitly propagating scheduling-induced demand downstream. However, the VS policy itself remains heuristic and cannot adapt through experience.

The heuristic VS is myopic by construction. It scores each candidate assignment by the immediate reduction in waiting and overflow time and thereby cannot trade a small immediate delay against a better future queue state or improve its policy through experience. Replacing it with a learning-based VS is thus a natural extension, but doing so qualitatively changes the overall EHTS operation from a single-agent learning setting, in which only the downstream ED learns, to a dual-agent learning problem in which both VS and ED policies learn. The preceding hierarchical framework \cite{yao2026hierarchical} found off-policy dispatchers (e.g., TD3) to outperform the on-policy dispatcher (e.g., proximal policy optimisation, PPO)  when the upstream scheduler was heuristic. Once the VS policy is also learned and evolves during training, however, the conditions supporting such replay-based learning change. As the VS policy changes, the charging and refuelling demand presented to the downstream ED also changes, so ED transitions collected under earlier VS policies may become progressively less representative of those induced by the current policy. This introduces a form of learning non-stationarity familiar from multi-agent reinforcement learning \cite{foerster2017stabilising, hernandez2019survey}, although, as shown later, the physical dependence in the present EHTS is directional from VS to ED rather than symmetric.

Beyond this interaction challenge, the two decision layers in the dual-agent setting pose fundamentally different learning problems. The VS action is combinatorial, taking the form of separate priority permutations over variable-size candidate sets of EVs and HVs, with the number of possible scheduling actions growing factorially with the sizes of the two candidate sets. By contrast, the ED action is a low-dimensional continuous control vector governing storage operation and PV allocation. This heterogeneity makes a common action representation or learning algorithm unnatural for both layers. The VS policy must represent and explore priority permutations without enumerating the factorial action space, while preserving a feasible mapping from the learned ranking to physical vehicle service. The ED policy, in contrast, must optimise continuous energy-flow decisions. The resulting dual-agent problem therefore requires not only coordination between the two decision layers, but also learning architectures suited to their fundamentally different decision structures.

This paper therefore addresses the following question: how can both vehicle scheduling and multi-energy dispatch be learned in a sequentially coupled EHTS while accounting for their heterogeneous decision structures and the dependence of downstream demand on scheduling decisions? The main contributions are as follows:

1) A dual-agent formulation is developed for joint VS and ED in EHTSs, enabling heterogeneous decision-making in both layers while preserving the directional VS$\rightarrow$ED operational dependence.

2) A heterogeneous RL architecture is developed for the different decision structures of the two layers. PPO with a Plackett--Luce (PL) policy learns EV and HV priority permutations, which are mapped to feasible pile assignments through deterministic greedy allocation; while TD3 controls battery and hydrogen-tank operation together with PV allocation.

3) A three-stage training mechanism is developed for the sequential dual-agent learning process. Layer-wise pretraining is followed by concurrent fine-tuning, where reduced ED replay capacity and update intensity mitigate the influence of older experience as the upstream VS policy evolves.

4) Extensive evaluations assess the proposed approach under varying operating and infrastructure conditions. The tests cover different PV penetration levels, vehicle arrival intensities, and charger/refueller capacities, examining the generalisation of the proposed framework beyond its training configuration.

The remainder of this paper is organised as follows. Section~\ref{sec:2} introduces the EHTS operational model, Section~\ref{sec:3} presents the dual-agent framework, Section~\ref{sec:4} reports the experimental evaluation, and Section~\ref{sec:5} concludes the paper.

%==============================
\section{EHTS Operational Model}
\label{sec:2}

\begin{figure}[t]
\centering
\includegraphics[width=8.5cm, trim=2 2 2 2
, clip]{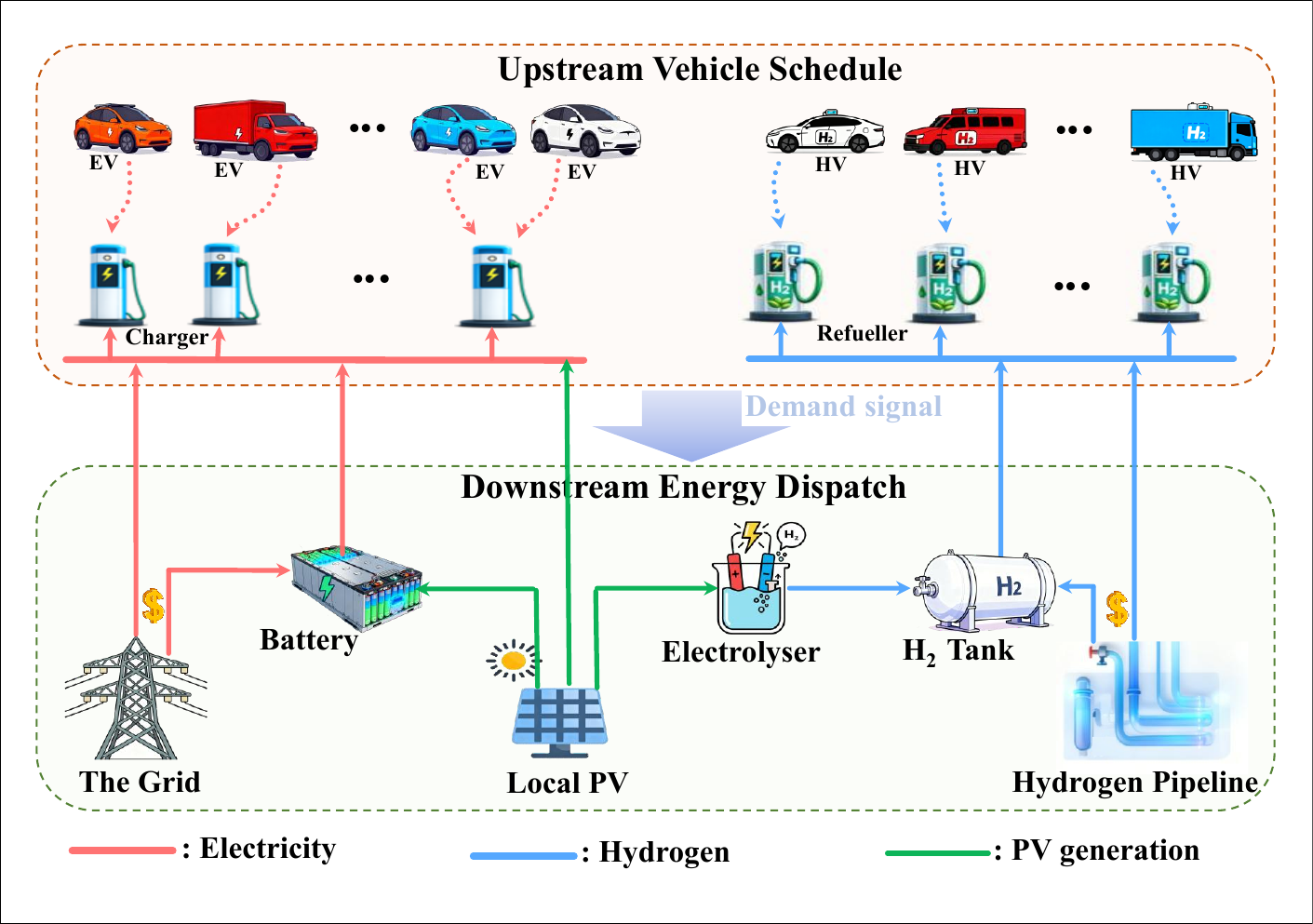}
% \figorbox{Fig1.pdf}{0.95\columnwidth}{3.2cm}
\captionsetup{font=small,skip=-5pt}
\caption{EHTS with vehicle scheduling and energy dispatch}
\label{fig:1}
\end{figure}

The studied EHTS, illustrated in Fig.~\ref{fig:1}, integrates multiple chargers and refuellers serving dynamically arriving EVs and HVs, together with PV generation, battery storage, an electrolyser, a hydrogen tank, grid connection, and hydrogen-pipeline supply. Operation is sequential within each interval of duration $\Delta t$: The physical station model follows \cite{yao2026hierarchical}, while the operational formulation required for the proposed dual-agent RL framework is stated below.

\subsection{Vehicle Scheduling Model} \label{sec:2-1}

At the beginning of interval $t$, newly arriving vehicles join those carried over from previous intervals to form the EV and HV waiting sets
$\mathcal{S}^{wt,ev}_t$ and $\mathcal{S}^{wt,hv}_t$, respectively. Each waiting EV $i$ is characterised by its remaining charging demand $E^{ev}_{i,t}$ (kWh) and accumulated waiting time $w^{ev}_{i,t}$ (h); similarly, $E^{hv}_{j,t}$ (Nm$^3$) and $w^{hv}_{j,t}$ (h) describe the remaining refuelling demand and waiting time of HV $j$. The station is assumed to contain $N^{ec}$ identical chargers at rated power $P^{ev,rt}$ (kW) and $N^{hr}$ identical refuellers at rated hydrogen flow $P^{hv,rt}$ (Nm$^3$/h).

Vehicle service follows two operational rules. First, service is non-preemptive: once connected to a charger or refueller, a vehicle remains in service until its current demand is satisfied, even if the service spans multiple decision intervals. 
% Vehicle service follows two operational rules. First, service is non-preemptive: once connected to a charger or refueller, a vehicle remains assigned to the same pile until its current demand is satisfied. If the service extends beyond the current decision interval, the unfinished service and the corresponding pile occupancy are carried forward to the subsequent interval(s).
Second, multiple vehicles may be served sequentially by the same pile within one decision interval if sufficient time remains after the preceding service is completed. Hence, VS determines not only which vehicles are served, but also their service priority and pile allocation.

\textbf{Non-Preemptive Service:} Let $c\in\{ev,hv\}$ denote the vehicle type. The amount of charging electricity or hydrogen delivered to vehicle $k$ during interval $t$ is
\begin{equation}
\small
e^c_{k,t}=
\begin{cases}
0, & \text{if not assigned},\\
\min\left(E^c_{k,t},P^{c,rt}\eta^c\Delta t\right),
& \text{if already assigned},\\
\min\left(E^c_{k,t},
P^{c,rt}\eta^c(\Delta t-\tau^{c,s}_{k,p,t})\right),
& \text{if newly assigned},
\end{cases}
\label{eq:1}
\end{equation}
where $\eta^c$ is the effective pile-to-vehicle service efficiency for vehicle type $c$ and $\tau^{c,s}_{k,p,t}
\in[0,\Delta t)$ is the service-start offset of vehicle $k$ on its assigned
pile $p$ within interval $t$. The remaining service demand then evolves as
\begin{equation}
\small
E^c_{k,t+1}=E^c_{k,t}-e^c_{k,t},
\label{eq:2}
\end{equation}
while the accumulated waiting time is updated as
\begin{equation}
\small
w^c_{k,t+1}=
\begin{cases}
w^c_{k,t}+\Delta t, & \text{if not assigned},\\
w^c_{k,t}, & \text{if already assigned},\\
w^c_{k,t}+\tau^{c,s}_{k,p,t}, & \text{if newly assigned}.
\end{cases}
\label{eq:3}
\end{equation}

\textbf{Within-Interval Sequential Assignment:} 
For an assigned vehicle $k$, its service duration on pile $p$ during interval $t$ is
\begin{equation}
\small
\tau^c_{k,p,t}=
\begin{cases}
0, & \text{if not assigned},\\
\min\left(
\frac{E^c_{k,t}}{P^{c,rt}\eta^c},
\Delta t
\right),
& \text{if already assigned},\\
\min\left(
\frac{E^c_{k,t}}{P^{c,rt}\eta^c},
\Delta t-\tau^{c,s}_{k,p,t}
\right),
& \text{if newly assigned}.
\end{cases}
\label{eq:4}
\end{equation}
After serving vehicle $k$, the remaining available service time of pile $p$
within the current interval is
\begin{equation}
\small
\tau^{c,rm}_{k,p,t}
=
\Delta t-\tau^{c,s}_{k,p,t}-\tau^c_{k,p,t}.
\label{eq:5}
\end{equation}
If $\tau^{c,rm}_{k,p,t}>0$, the residual time can be allocated to
the next vehicle according to the scheduling priority. The procedure is
repeated until no additional vehicle can be accommodated within the current
interval. Here, the transition time between consecutive vehicles is neglected. Accordingly, the cumulative occupied time of pile $p$ during interval $t$
is
\begin{equation}
\small
\tau^c_{p,t}
=
\sum_{k\in\mathcal{S}^{wt,c}_t}
\tau^c_{k,p,t},
\qquad
0\leq\tau^c_{p,t}\leq\Delta t.
\label{eq:6}
\end{equation}

\textbf{Scheduling-Induced Energy Demand:}
The VS determines the utilisation of each charger and refueller and thereby generates the aggregate electricity and hydrogen demands passed to the downstream ED layer. Based on the cumulative pile occupation time in \eqref{eq:6}, the aggregate electricity demand of the charger cluster is
\begin{equation}
\small
P^{ev}_t = \sum_{p=1}^{N^{ec}}
\frac{\tau^{ev}_{p,t}}{\Delta t}
P^{ev,rt},
H^{hv}_t = \sum_{p=1}^{N^{hr}}
\tau^{hv}_{p,t}
P^{hv,rt},
\label{eq:7}
\end{equation}
where $P^{ev}_t$ (kW) and $H^{hv}_t$ (Nm$^3$) denote the aggregate electricity and hydrogen demand of the charger and refueller clusters during interval $t$. The resulting $P^{ev}_t$ and $H^{hv}_t$
constitute the demand inputs to the downstream multi-energy dispatch model.

\subsection{Multi-energy Dispatch Model} \label{sec:2-2}
The ED layer coordinates energy flow among PV generation, battery storage, electrolyser, hydrogen tank, and external energy supplies to satisfy the two demands ($P^{ev}_t$ and $H^{hv}_t$). The grid and hydrogen pipeline are both modelled as import-only external supplies with sufficient capacity to meet any residual electricity and hydrogen demand, respectively. The available PV generation $P^{pv}_t$ is treated as a renewable input here, while its generation model follows \cite{yao2026hierarchical} and is omitted here for brevity.

\textbf{Battery Storage:}
The battery can be charged by PV generation or grid electricity and
discharged to support the EV charging demand. Its state of charge (SOC)
evolves as
\begin{equation}
\small
SOC^{bt}_{t+1}
=
SOC^{bt}_t
-\frac{P^{bt,ev}_t\Delta t}
{E^{bt,cp}\eta^{bt,dc}}
+\frac{
\left(P^{pv,bt}_t+P^{gd,bt}_t\right)
\Delta t\eta^{bt,ch}}
{E^{bt,cp}},
\label{eq:9}
\end{equation}
where $E^{bt,cp}$ (kWh) is the battery capacity,
$P^{bt,ev}_t$ (kW) is the battery power supplied to the charger cluster for EV demand, while $P^{pv,bt}_t$ and $P^{gd,bt}_t$ (kW) denote the battery charging
powers supplied by PV generation and the grid, respectively.
$\eta^{bt,ch}$ and $\eta^{bt,dc}$ are the battery charging and discharging efficiencies. The battery operation is subject to its rated power limits:
\begin{equation}
\small
0\leq P^{bt,ev}_t\leq P^{bt,max}, \qquad
P^{pv,bt}_t+P^{gd,bt}_t\leq P^{bt,max},
\label{eq:10}
\end{equation}
where $P^{bt,max}$ is the maximum battery charging/discharging power. To ensure that the battery SOC remains within its allowable range, the discharging and charging powers must further satisfy
\begin{equation}
\small
P^{bt,ev}_t
\leq
\frac{
\left(SOC^{bt}_t-SOC^{bt,min}\right)
E^{bt,cp}\eta^{bt,dc}}
{\Delta t},
\label{eq:11}
\end{equation}
\begin{equation}
\small
P^{pv,bt}_t+P^{gd,bt}_t
\leq
\frac{
\left(SOC^{bt,max}-SOC^{bt}_t\right)
E^{bt,cp}}
{\eta^{bt,ch}\Delta t},
\label{eq:12}
\end{equation}
where $SOC^{bt,min}$ and $SOC^{bt,max}$ are the minimum and maximum allowable battery SOC levels, respectively.

\textbf{Electrolyser and Hydrogen Tank:}
Surplus PV generation can be converted into hydrogen through the electrolyser. The hydrogen produced during interval $t$ is
% In the present model, on-site electrolysis is not assumed to be the primary hydrogen supply route for the vehicle fleet. Instead, the electrolyser provides a flexible pathway for utilising locally available surplus PV generation, while bulk hydrogen demand can also be supplied through the external hydrogen pipeline. This modelling choice reflects a future integrated energy-hub setting in which local renewable hydrogen production complements, rather than replaces, network-based hydrogen supply. Detailed electrolyser part-load efficiency, compression, purification, and high-pressure dispensing processes are outside the scope of the present operational model and are represented through aggregate conversion and supply parameters.
\begin{equation}
\small
H^{el,tk}_t = \frac{P^{pv,el}_t\Delta t}{\eta^{el}},
\label{eq:13}
\end{equation}
where $P^{pv,el}_t$ (kW) is the PV power supplied to the electrolyser and $\eta^{el}$ (kWh/Nm$^3$) denotes the electricity required to generate one unit of hydrogen.

The hydrogen tank stores hydrogen produced by the electrolyser or supplied
by the hydrogen pipeline and can discharge to support HV refuelling
demand. Its SOC evolves as
\begin{equation}
\small
SOC^{tk}_{t+1}
=
SOC^{tk}_t
-\frac{H^{tk,hv}_t}
{H^{tk,cp}\eta^{tk,dc}}
+\frac{
\left(H^{el,tk}_t+H^{pl,tk}_t\right)
\eta^{tk,ch}}
{H^{tk,cp}},
\label{eq:14}
\end{equation}
where $H^{tk,cp}$ (Nm$^3$) is the hydrogen tank capacity,
$H^{tk,hv}_t$ (Nm$^3$) denotes hydrogen supplied from the tank to the refueller cluster for HV demand, and $H^{pl,tk}_t$ (Nm$^3$) denotes hydrogen supplied from the pipeline to the tank. $\eta^{tk,ch}$ and $\eta^{tk,dc}$ are
the tank charging and discharging efficiencies. The hydrogen-tank operation is subject to its charging/discharging limits:
\begin{equation}
\small
0\leq H^{tk,hv}_t\leq H^{tk,max},
0\leq H^{el,tk}_t+H^{pl,tk}_t\leq H^{tk,max},
\label{eq:15}
\end{equation}
where $H^{tk,max}$ denotes the maximum hydrogen charging/discharging amount within one decision interval.
The hydrogen flows must further satisfy
\begin{equation}
\small
H^{tk,hv}_t
\leq
\left(SOC^{tk}_t-SOC^{tk,min}\right)
H^{tk,cp}\eta^{tk,dc},
\label{eq:16}
\end{equation}
\begin{equation}
\small
H^{el,tk}_t+H^{pl,tk}_t
\leq
\frac{
\left(SOC^{tk,max}-SOC^{tk}_t\right)
H^{tk,cp}}
{\eta^{tk,ch}},
\label{eq:17}
\end{equation}
where $SOC^{tk,min}$ and $SOC^{tk,max}$ denote the minimum and maximum allowable hydrogen-tank SOC levels, respectively.

\textbf{Multi-Energy Balance:} Available PV generation $P_t^{pv}$ (kW) is prioritised to supply the EV charging
demand, such that
\begin{equation}
\small
P_t^{pv,ev}
=
\min\left(P_t^{pv},P_t^{ev}\right),
P_t^{pv,sur}
=
\max\left(P_t^{pv}-P_t^{ev},0\right),
\label{eq:17-1}
\end{equation}
where $P_t^{pv,ev}$ (kW) denotes the PV power supplied to the charger
cluster and $P_t^{pv,sur}$ (kW) denotes the resulting surplus PV power.
The latter can be allocated to battery charging and hydrogen
production through the electrolyser, while any remaining PV power is curtailed. This operating structure prioritises
the direct use of locally available PV generation for the immediate
EV charging load before storage or energy conversion. In the present EHTS configuration, the electrolyser is modelled specifically as a surplus-renewable-to-hydrogen pathway to utilise excess PV generation and reduce renewable curtailment; grid-powered electrolysis is therefore outside the considered operating scope. The PV power balance is
\begin{equation}
\small
P^{pv}_t =
P^{pv,ev}_t + P^{pv,bt}_t + P^{pv,el}_t + P^{pv,cur}_t,
\label{eq:18}
\end{equation}
where $P^{pv,cur}_t$ (kW) denotes the curtailed PV power.

The scheduling-induced EV charging demand in \eqref{eq:7} is satisfied by PV generation, battery discharge, and grid import:
\begin{equation}
\small
P^{ev}_t =
P^{pv,ev}_t + P^{bt,ev}_t + P^{gd,ev}_t,
\label{eq:19}
\end{equation}
where $P^{gd,ev}_t$ (kW) denotes the grid power supplied directly to
the charger cluster.

Similarly, the scheduling-induced hydrogen demand in \eqref{eq:7} is
satisfied by hydrogen-tank discharge and pipeline supply:
\begin{equation}
\small
H^{hv}_t = H^{tk,hv}_t + H^{pl,hv}_t,
\label{eq:20}
\end{equation}
where $H^{pl,hv}_t$ (Nm$^3$) denotes hydrogen supplied directly from
the pipeline to the refueller cluster for HV demand.

% The grid and hydrogen pipeline are assumed to provide sufficient external backup, such that any electricity or hydrogen demand not supplied by local resources can be satisfied externally. The grid connection is import-only, and surplus PV generation that cannot be consumed by the charger cluster, stored in the battery, or converted into hydrogen is curtailed. 
Therefore, the ED layer determines how the electricity and hydrogen demands generated by the upstream VS layer are supplied through coordinated use of local storage, renewable generation, and external energy sources.

%=============================
\section{Dual-Agent Reinforcement Learning}
\label{sec:3}
Fig. 2 presents the proposed heterogeneous dual-agent framework and the sequential interaction between the two agents. The vehicle-scheduling (VS) agent uses PPO with a PL policy to generate vehicle-priority permutations, and a deterministic greedy allocator converts these priorities into feasible service assignments. The energy-dispatch (ED) agent uses TD3 to determine continuous storage and PV-allocation actions conditioned on the resulting demand. A physical mapping then converts the ED action into feasible energy flows. Both agents adopt actor--critic
architectures but operate with separate states, actions, rewards, and learning updates. 

% =========================================
\subsection{Overall Framework and Sequential Coupling}
\label{sec:3-1}

\begin{figure}[t]
\centering
\includegraphics[width=9cm, trim=2 2 2 2
, clip]{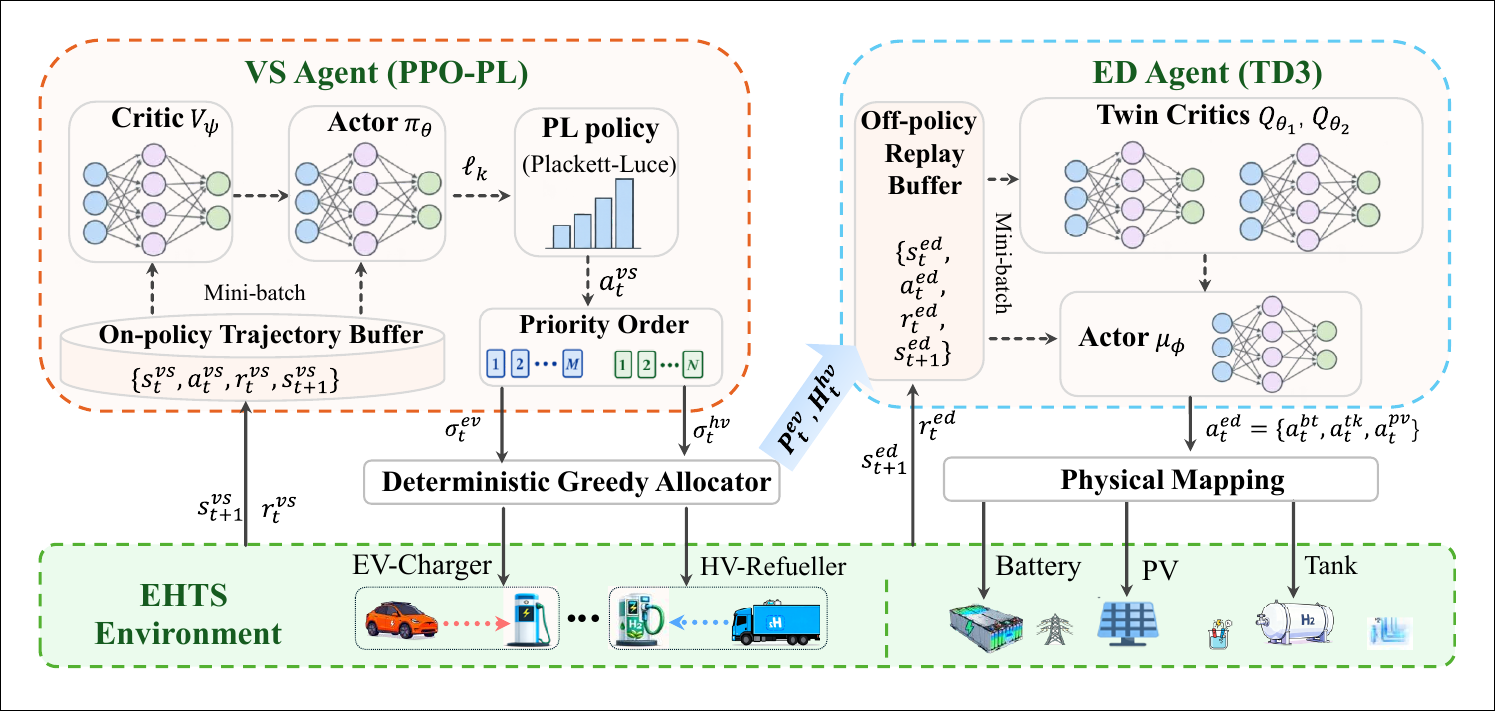}
% \figorbox{Fig1.pdf}{0.95\columnwidth}{3.2cm}
\captionsetup{font=small,skip=5pt}
\caption{Heterogeneous dual-agent reinforcement learning}
\label{fig:2}
\end{figure}

To maintain a bounded scheduling dimension, at time $t$, at most $M$ EVs and $N$ HVs are selected from the waiting sets $\mathcal{S}^{wt,ev}_t$ and $\mathcal{S}^{wt,hv}_t$ as scheduling candidates, respectively.
Within each vehicle class, candidates are selected in descending order of accumulated waiting time, with smaller remaining service demand used as the tie-breaker. When fewer than $M$ EVs or $N$ HVs are available in $\mathcal{S}^{wt,ev}_t$ or $\mathcal{S}^{wt,hv}_t$, the remaining candidate slots are zero-padded and masked to exclude them from the later ranking process. Rather than directly learning vehicle--pile assignments, the VS agent learns only vehicle-service priorities. Direct assignment would introduce a huge combinatorial action space and couple the learned policy to a fixed number of chargers and refuellers. Separating priority learning from pile allocation therefore reduces the learning complexity and improves portability across different service-infrastructure configurations. The PPO-based VS agent then receives the transport-side state $\textbf{s}_t^{vs}$, and its actor network produces a policy logit $\ell_k$ for each valid EV and HV candidate. These logits parameterize separate Plackett--Luce (PL) distributions \cite{plackett1975analysis} for the EV and HV candidate sets, from which the priority permutations $\sigma_t^{ev}$ and $\sigma_t^{hv}$ are sequentially sampled without replacement. 
Let $\mathcal{C}_t^c$ denote the set of valid candidates of class $c\in\{ev,hv\}$, where
$|\mathcal{C}_t^{ev}|\leq M$ and $|\mathcal{C}_t^{hv}|\leq N$. For either vehicle class,
the probability of a permutation
$\sigma^c$ is
\begin{equation}
\small
\pi_{\theta}
\left(\sigma^c\mid s_t^{vs}\right)
=
\prod_{r=1}^{|\mathcal{C}_t^c|}
\frac{
\exp\!\left(\ell_{\sigma^c(r)}\right)
}{
\displaystyle
\sum_{j\in\mathcal{U}_r^c}
\exp(\ell_j)
},
\label{eq:21}
\end{equation}
where $\mathcal{U}_r^c\subseteq\mathcal{C}_t^c$ denotes the set of unselected candidates of class $c$ before the $r$-th selection. The sampled permutations determine the order in which candidate vehicles are considered for service. A deterministic greedy allocator then maps these priority orders to feasible pile assignments. For each
vehicle class $c\in\{ev,hv\}$, vehicles are processed sequentially according to $\sigma_t^c$, and each vehicle is assigned to the compatible pile with the earliest feasible service
start. This procedure continues along the priority order until all candidate vehicles have been assigned or no additional service can be accommodated within the current interval.

The resulting assignments determine the within-interval utilisation $\tau_{p,t}^{c}$ of each pile according to (\ref{eq:6}). The corresponding electricity and hydrogen demands $P_t^{ev}$ and $H_t^{hv}$ are then calculated from the EV-charger and HV-refueller utilisation according to (\ref{eq:7}). Since the priority permutations and pile assignments are determined at the start of time $t$, $P_t^{ev}$ and $H_t^{hv}$ are known before the ED decision and represent the scheduling-induced demands to be supplied over the upcoming interval $[t,t+1)$.

The TD3-based ED agent subsequently receives the energy-side state $\textbf{s}_t^{ed}$, which incorporates the scheduling-induced demands $P_t^{ev}$ and $H_t^{hv}$ together with the current energy-system information. Based on $\textbf{s}_t^{ed}$, the TD3 actor generates a continuous dispatch action $\textbf{a}_t^{ed}$, which is mapped to feasible battery and hydrogen-tank decisions and PV allocation for the interval $[t,t+1)$. Accordingly, the within-interval sequential coupling between the two agents can be summarized as
\begin{equation}
\small
   \textbf{s}_t^{vs} 
   \rightarrow \textbf{a}_t^{vs}
    \rightarrow \left(P_t^{ev},H_t^{hv}\right)
    \rightarrow  \textbf{s}_t^{ed}
    \rightarrow \textbf{a}_t^{ed}.
    \label{eq:22}
\end{equation}

After the VS and ED decisions are determined, the corresponding vehicle
service and energy dispatch are executed over $[t,t+1)$, and the
EHTS transitions to the next state at $t+1$. Therefore, the coupling
is directional: the VS decision determines the electricity and
hydrogen demands handled by the ED agent, whereas the ED decision
does not alter the vehicle service already determined for
the same interval.

\subsection{Vehicle Scheduling Agent}
\label{sec:3-2}
\textbf{State:} The VS agent learns the service priorities of the selected EV and HV candidates according to the current transport-side state
\begin{equation}
\small
\textbf{s}_t^{vs}
=
\left[
\mathbf{x}_t^{ev},
\mathbf{x}_t^{hv},
\mathbf{z}_t^{vs}
\right],
\label{eq:23}
\end{equation}
where $\mathbf{x}_t^{ev}\in\mathbb{R}^{M\times 4}$ and
$\mathbf{x}_t^{hv}\in\mathbb{R}^{N\times 4}$ contain the candidate-level features of the selected EVs and HVs, respectively; and $\mathbf{z}_t^{vs}$ collects the aggregate transport-side information. For each vehicle candidate $k$ of vehicle class $c\in\{ev,hv\}$, the candidate feature vector is
\begin{equation}
\small
\mathbf{x}_{k,t}^{c}
= \left[ E_{k,t}^{c}, w_{k,t}^{c},
d_{k,t}^{c}, \tau_{k,t}^{c}
\right],
\label{eq:24}
\end{equation}
where $d_{k,t}^{c}$ denotes the remaining waiting slack
before vehicle $k$ reaches the prescribed waiting-time limit, while $\tau_{k,t}^{c}$ is the service duration implied by the vehicle's remaining demand and the rated service rate. 

The aggregate transport-side vector is defined as
\begin{equation}
\small
\mathbf{z}_t^{vs}
=
\left[
\mathbf{z}_t^{pile},
\mathbf{z}_t^{que},
\mathbf{z}_t^{un},
\mathbf{z}_t^{time},
\right],
\label{eq:25}
\end{equation}

where $\mathbf{z}_t^{pile}$ contains the residual occupied time of all EV chargers and HV refuellers; $\mathbf{z}_t^{que}$ contains the numbers of waiting and in-service EVs and HVs; $\mathbf{z}_t^{ua}$ contains the aggregate unassigned EV charging and HV refuelling demands within $\mathcal{S}^{wt,ev}_t$ and $\mathcal{S}^{wt,hv}_t$; and $\mathbf{z}_t^{time}$ contains the current hour-of-day and day-of-week information. A fixed number of pile slots is retained in the VS state, and unavailable chargers/refuellers are zero-padded so that the state dimension remains unchanged across infrastructure configurations. All state variables are normalised to $[0,1]$ using min--max normalisation.

\textbf{Action:} Given $\mathbf{s}_t^{vs}$, the actor network produces ranking
logits for all valid candidates. As introduced in Section~\ref{sec:3-1}, these logits parameterize the PL distributions from which the EV and HV priority permutations are sampled. Accordingly, the VS action is defined as
\begin{equation}
\small
\mathbf{a}_t^{vs}
= \left(
\sigma_t^{ev},\sigma_t^{hv}
\right),
\qquad
\sigma_t^{c}\in\Pi(\mathcal{C}_t^{c}),
\quad c\in\{ev,hv\},
\label{eq:26}
\end{equation}
where $\Pi(\mathcal{C}_t^{c})$ denotes the set of all permutations of the valid candidate set $\mathcal{C}_t^{c}$. Thus, the PPO agent learns vehicle service priorities rather than vehicle--pile assignments. The sampled permutations are then mapped to feasible pile assignments by the deterministic greedy allocator described in Section~\ref{sec:3-1}.

\textbf{Reward:} The VS agent is designed to reduce vehicle waiting and excessive waiting beyond the prescribed thresholds $w^{c,\max}$, $c\in\{ev,hv\}$. Its reward at interval $t$ is defined as
\begin{equation}
\small
r_t^{vs}
=
-\left(
\omega^{wt} C_t^{wt}
+
\omega^{ot} C_t^{ot}
\right),
\label{eq:27}
\end{equation}
where
\begin{equation}
\small
C_t^{wt}
=
\sum_{c\in\{ev,hv\}}
\sum_{k\in\mathcal{S}^{wt,c}_t}
\left(
w^c_{k,t+1}-w^c_{k,t}
\right),
\label{eq:27-1}
\end{equation}
\begin{equation}
\small
C_t^{ot}
=
\sum_{c\in\{ev,hv\}}
\sum_{k\in\mathcal{S}^{wt,c}_t}
\max\left(
0,w^c_{k,t+1}-w^{c,\max}
\right).
\label{eq:27-2}
\end{equation}
Here, $C_t^{wt}$ (h) denotes the total waiting-time increment of all vehicles during interval $t$, while $C_t^{ot}$ (h) denotes the corresponding overflow time beyond the prescribed class-specific waiting-time thresholds. The coefficients $\omega^{wt}$ and $\omega^{ot}$ are the corresponding penalty weights.

\textbf{Optimisation:} The VS policy is optimised using PPO \cite{schulman2017proximal}. Conditional on the transport state, the EV and HV permutations are sampled independently, so the joint action log-likelihood is the sum of their PL log-likelihoods defined in (\ref{eq:21}). The PPO likelihood ratio is computed for the complete sampled permutations under the current and behaviour policies. Invalid candidate slots are excluded, and an empty candidate set contributes zero log-likelihood. Policy updates use the clipped PPO objective with advantages computed using generalized advantage estimation.

% \textbf{Optimisation:} The PL ranking policy is optimised using PPO \cite{schulman2017proximal}. Let $\pi_\theta$ denote the PPO actor parameterized by $\theta$ and $V_\psi$ denote the corresponding critic parameterized by $\psi$. Since $\mathbf{a}_t^{vs}$ consists of complete EV and HV permutations, its log-likelihood is obtained from the corresponding PL distributions. The probability ratio between the current and old policies is
% \begin{equation}
% \small
% \rho_t(\theta)
% = \exp\!\left[
% \log\pi_{\theta}
% \left(\mathbf{a}_t^{vs}\mid \mathbf{s}_t^{vs}\right)
% -
% \log\pi_{\theta_{old}}
% \left(\mathbf{a}_t^{vs}\mid \mathbf{s}_t^{vs}\right)
% \right],
% \label{eq:28}
% \end{equation}
% where $\theta_{old}$ denotes the actor parameters before the
% current PPO update. The actor network is updated by maximizing the clipped objective
% \begin{equation}
% \small
% \mathcal{J}_{PPO}(\theta)
% =
% \mathbb{E}_t
% \left[
% \min
% \left(
% \rho_t(\theta)\hat{A}_t,\,
% \operatorname{clip}
% \left(\rho_t(\theta),1-\epsilon,1+\epsilon\right)\hat{A}_t
% \right)
% \right],
% \label{eq:29}
% \end{equation}
% where $\hat{A}_t$ is the advantage estimate and $\epsilon$ is the PPO clipping parameter. The critic $V_\psi(\mathbf{s}_t^{vs})$ estimates the state value, and $\hat{A}_t$ is computed using generalized advantage estimation (GAE) \cite{schulman2017proximal}.

\subsection{Energy Dispatch Agent}
\label{sec:3-3}
The ED agent adopts the TD3-based dispatch design in \cite{yao2026hierarchical}, with its demand inputs now generated by the learned VS policy. The dispatch state, action, and reward are summarised below to make the interaction between the two agents explicit. The principal change considered in this work is the evolving upstream scheduling policy and its implications for downstream training.

\textbf{State:} The ED state combines the current operating condition with historical system information and is expressed as
\begin{equation}
\small
\mathbf{s}_t^{ed}
=
\left[
\mathbf{z}_t^{cur},
\mathbf{z}_t^{his}
\right],
\label{eq:30}
\end{equation}
where $\mathbf{z}_t^{cur}$ contains the current unmet power $P_t^{um}$, hydrogen refuelling demand $H_t^{hv}$, electricity and hydrogen prices $p_t^{elc}$ and $p_t^{h2}$, battery and hydrogen-tank states of charge $SOC_t^{bt}$ and $SOC_t^{tk}$, the unassigned EV charging and HV refuelling demands $P_t^{ev,ua}$ and $H_t^{hv,ua}$ (corresponding to $\mathbf{z}_t^{ua}$ defined in the VS state), and the hour-of-day and day-of-week information. The historical component contains the electricity price, hydrogen price, unmet power, and hydrogen refuelling demand
\begin{equation}
\small
\mathbf{z}_t^{his}
=\left[
\mathbf{p}_{\mathcal{T}_t}^{elc},
\mathbf{p}_{\mathcal{T}_t}^{h2},
\mathbf{P}_{\mathcal{T}_t}^{um},
\mathbf{H}_{\mathcal{T}_t}^{hv}
\right],
\mathcal{T}_t
=\{t-4l\mid l=1,\ldots,L\},
\label{eq:31}
\end{equation}
where $L$ is the number of historical observations. Following \cite{yao2026hierarchical}, $\Delta t=0.25$ and $L=6$ with a stride of four is adopted under the same decision interval, such that $\mathcal{T}_t=\{t-4,t-8,\ldots,t-24\}$ covers the previous six hours.

\textbf{Action:} The ED agent determines the charging/discharging decisions of the battery and hydrogen tank, together with the allocation of surplus PV generation. The action is defined as
\begin{equation}
\small
\mathbf{a}_t^{ed}
=
\left[
a_t^{bt},
a_t^{tk},
a_t^{pv}
\right],
\qquad
\mathbf{a}_t^{ed}\in[-1,1]^3,
\label{eq:32}
\end{equation}
where positive (negative) values of $a_t^{bt}$ and $a_t^{tk}$ represent discharging (charging) actions of battery and hydrogen tank, respectively. The storage actions are then mapped to the corresponding physical quantities as
\begin{equation}
\small
a_t^{bt}P^{bt,\max}
=
\begin{cases}
P_t^{bt,ev}, & a_t^{bt}\geq0,\\
-\left(P_t^{pv,bt}+P_t^{gd,bt}\right), & a_t^{bt}<0,
\end{cases}
\label{eq:33}
\end{equation}

\begin{equation}
\small
a_t^{tk}H^{tk,\max}
=
\begin{cases}
H_t^{tk,hv}, & a_t^{tk}\geq0,\\
-\left(H_t^{el,tk}+H_t^{pl,tk}\right), & a_t^{tk}<0,
\end{cases}
\label{eq:34}
\end{equation}
subject to the storage operating constraints in (\ref{eq:10})--(\ref{eq:12}) and (\ref{eq:15})--(\ref{eq:17}), respectively.

The PV action is mapped to $\alpha_t^{pv}=(a_t^{pv}+1)/2\in[0,1]$, which allocates the surplus PV power defined in (\ref{eq:17-1}) as
\begin{equation}
\small
P_t^{pv,bt}
=
\alpha_t^{pv}P_t^{pv,sur},
\qquad
P_t^{pv,el}
=
(1-\alpha_t^{pv})P_t^{pv,sur},
\label{eq:35}
\end{equation}
subject to the corresponding operating constraints in (\ref{eq:10})--(\ref{eq:17}).

\textbf{Reward:} The ED reward measures the reduction in operating cost relative to a baseline operation \cite{yao2025unified}:
\begin{equation}
\small
r_t^{ed}
=
\left(C_t^{gd,base}-C_t^{gd}\right)
+
\left(C_t^{pl,base}-C_t^{pl}\right)
+
\left(C_t^{pv,base}-C_t^{pv}\right),
\label{eq:36}
\end{equation}
where
\begin{equation}
\small
\begin{cases}
C_t^{gd}
=
p_t^{elc}
\left(P_t^{gd,ev}+P_t^{gd,bt}\right)\Delta t,\\
C_t^{pl}
=
p_t^{h2}
\left(H_t^{pl,hv}+H_t^{pl,tk}\right),\\
C_t^{pv}
=
p_t^{elc}P_t^{pv,cur}\Delta t.
\end{cases}
\label{eq:35}
\end{equation}
Here, $C_t^{gd}$, $C_t^{pl}$, and $C_t^{pv}$ denote the grid-purchase, hydrogen-pipeline-purchase, and PV-curtailment costs, respectively. The PV-curtailment cost represents the opportunity cost of renewable generation that could otherwise offset grid purchases. The corresponding baseline costs $C_t^{*,base}$ are evaluated without battery and hydrogen-tank utilisation: unmet electricity demand is supplied entirely by the grid, hydrogen refuelling demand entirely by the hydrogen pipeline, and all surplus PV generation is curtailed.

\textbf{Optimisation:} The ED policy is optimised using TD3 \cite{fujimoto2018addressing}, with twin critics, target-policy smoothing, delayed actor updates, and soft target-network updates. Completed ED transitions are sampled from the replay buffer, with each next state constructed after the subsequent VS decision generates its demand inputs. The stage-dependent replay capacity and update intensity are specified in Section \ref{sec:3-4}, and the training process is provided in Appendix \ref{app:1}.

% \textbf{Optimisation:} The ED policy is optimised using TD3 \cite{fujimoto2018addressing}. Let $\mu_\phi$ denote the TD3 actor parameterized by $\phi$, and let $Q_{\theta_1}$ and $Q_{\theta_2}$ denote the twin critics parameterized by $\theta_1$ and $\theta_2$, respectively. The critic target is defined as
% \begin{equation}
% \small
% y_t = r_t^{ed} + \gamma \min_{i=1,2} Q_{\theta_i'}
% \left( \mathbf{s}_{t+1}^{ed}, \mathbf{a}_{t+1}^{ed'} \right),
% \label{eq:38}
% \end{equation}
% where $\gamma$ is the discount factor, $\mathbf{a}_{t+1}^{ed'}$ is generated by the target actor with clipped smoothing noise, and $\theta_i'$ denotes the parameters of the $i$th target critic. The actor network is updated by maximizing the action value estimated by the first critic, equivalently by minimizing
% \begin{equation}
% \small
% \mathcal{L}_{\mu}(\phi)
% = - \mathbb{E}
% \left[ Q_{\theta_1}
% \left(
% \mathbf{s}_t^{ed},
% \mu_{\phi}(\mathbf{s}_t^{ed})
% \right) \right].
% \label{eq:39}
% \end{equation}
% Delayed actor updates and soft target-network updates are applied following the standard TD3 procedure.

\subsection{Algorithmic Design and Three-Stage Training}\label{sec:3-4} 
The two agents adopt different RL algorithms to match their distinct physical and decision characteristics. VS is a rapidly varying ranking problem, where stochastic arrivals and varying vehicle demands continuously reshape the candidate set. Accordingly, on-policy PPO is adopted to directly optimise the likelihood of the executed PL permutations under the current policy. In contrast, ED involves low-dimensional continuous control with inter-temporal storage dynamics and temporally correlated energy-side conditions. Off-policy TD3 is therefore employed to reuse historical experience through replay, while its twin-critic and delayed policy updates support robust continuous control.

Updates of the upstream VS policy can shift the demand distribution experienced by ED, making earlier ED transitions less representative of the current demand process. A three-stage mechanism is adopted to accommodate this variation. Stage~I pretrains PPO-PL with the fixed rule-based ED controller in \cite{yao2026hierarchical}; Stage~II freezes PPO-PL and pretrains TD3 under the resulting demand; and Stage~III updates both policies concurrently using their respective rewards. During Stage~III, the ED replay capacity is reduced to 10\% of its Stage-II value and the TD3 update intensity to one-third, with the aim of limiting reliance on earlier experience as VS evolves. 

Stage transitions are plateau-triggered rather than predetermined: performance is evaluated every 50 epochs by comparing two consecutive 100-epoch windows, and a stage advances after four consecutive reward checks without an improvement greater than $0.01$. The complete seudocode is provided in Appendix \ref{app:1}.
%==============================

\section{Experiments}
\label{sec:4}
This section evaluates the proposed framework in terms of algorithmic performance, training effectiveness, operational behaviour, and generalisation capability. The experiments consider an EHTS case study in Wales, UK, using hybrid annual datasets that combine real-world meteorological and electricity-price data with synthetically generated hydrogen-price and transport-demand profiles. Four year-long high-
demand datasets are constructed: one for training and three independently generated datasets for testing. All datasets use a 15-min resolution. The nominal EHTS contains five EV chargers and three hydrogen refuellers and the complete three-stage training procedure is conducted for at most 20,000 epochs. Five independently trained runs with random seeds 42--46 are used for each DRL configuration. Generalisation is further evaluated using medium- and low-demand profiles, as well as under unseen PV penetration levels and service-infrastructure conditions, as detailed in Section~\ref{sec:4-6}. The complete dataset construction, physical configuration, and DRL settings are provided in Appendix~\ref{app:1-1}.

\subsection{Vehicle-Service Policy Comparison}
\label{sec:4-1}
We first compare the proposed PPO-based VS policy with representative heuristic and DRL scheduling alternatives. For clarity in this comparison, the proposed PPO policy with the PL action parameterisation is denoted as PPO-PL, while PPO-score and TD3-score denote the corresponding score-based variants. The heuristic baselines are FIFO, a conventional queue-order rule, and SFGH, the best-performing heuristic scheduler identified in \cite{yao2026hierarchical}. The score-based variants emit one real-valued priority for each candidate and obtain the service order by sorting, rather than directly sampling a permutation. Comparing PPO-PL with PPO-score isolates the effect of action parameterisation, while comparing PPO-score with TD3-score isolates the effect of the learning algorithm. A corresponding TD3-PL variant is not considered because TD3 requires a continuous action representation, whereas the PL policy directly represents discrete priority permutations. The DRL scheduling variants are paired with TD3, while the heuristic schedulers are paired with the rule-based (RB) ED controller used in their reference configurations \cite{yao2026hierarchical}. Additional cross-layer pairings are examined separately in Section~\ref{sec:4-3}. Following the definitions in Section~\ref{sec:3-2}, $C^{wt}$ and $C^{ot}$ denote the cumulative waiting and overflow times over the evaluation horizon (one year), respectively. Since $\omega^{wt}=\omega^{ot}=1$ in all experiments, we use $C^{svc}\triangleq C^{wt}+C^{ot}$ as the overall vehicle-service metric. Table~\ref{tab:1} reports these metrics averaged over the three held-out test datasets. For each DRL method, the results are further averaged over five independent training runs using different random seeds; dataset-specific results are provided in Appendix~\ref{app:2}.

\begin{table}[t]
\caption{Vehicle-scheduling delays (h) averaged over three test sets and over five independent runs}
\label{tab:1}
\centering
\footnotesize
\setlength{\tabcolsep}{3.5pt}
\begin{tabular}{lrrr@{\hspace{10pt}}lrrr}
\toprule
\textbf{Method} & $C^{wt}$ & $C^{ot}$ & $C^{svc}$
& \textbf{Method} & $C^{wt}$ & $C^{ot}$ & $C^{svc}$ \\
\midrule
\textbf{PPO-PL + TD3}
& \textbf{7722 }& \textbf{905} & \textbf{8627}
& FIFO + RB
& 7928 & 981 & 8910 \\

TD3-score + TD3
& 7888 & 1174 & 9062
& SFGH + RB
& 7749 & 955 & 8704 \\

PPO-score + TD3
& 7876 & 950 & 8826
& & & & \\
\bottomrule
\multicolumn{8}{l}{\footnotesize
Discrepancies of up to 1 may occur due to rounding; same to tables below.}
\end{tabular}
\end{table}
\begin{figure}[t]
\centering
\includegraphics[width=9.0cm, height=3.5cm, trim=1 1 1 1, clip]{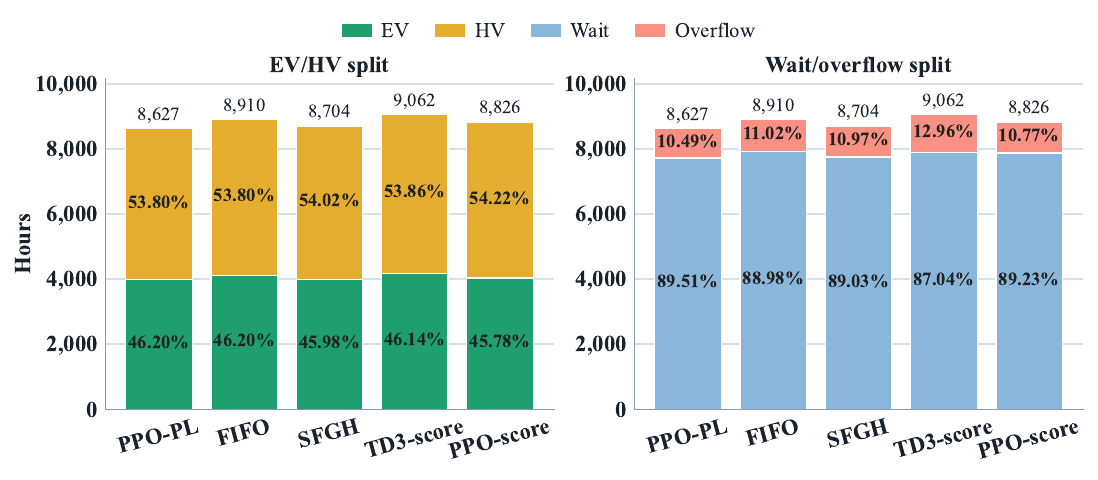}
% \figorbox{Fig1.pdf}{0.95\columnwidth}{3.2cm}
\captionsetup{font=small,skip=0pt}
\caption{Vehicle-service delay decomposition}
\label{fig:3}
\end{figure}
PPO-PL achieves the lowest $C^{wt}$, $C^{ot}$, and $C^{svc}$ among the compared methods, demonstrating that the learned scheduler can outperform heuristic scheduling by adapting its decisions to the evolving vehicle-service state. In particular, PPO-PL reduces $C^{ot}$ by 7.7\% relative to FIFO and by 5.2\% relative to SFGH. Despite this improvement, waiting time remains the dominant component for all methods in Fig.~\ref{fig:3}, accounting for 87--90\% of the combined service delay. Across all schedulers, EVs and HVs consistently account for approximately 46\% and 54\% of $C^{svc}$, respectively. This stable distribution also indicates that the improvement achieved by PPO-PL is not primarily driven by shifting delay from one vehicle class to the other.
\begin{table}[b]
\caption{ED policy comparison (k\pounds) under PPO-PL VS}
\label{tab:2}
\centering
\footnotesize
\setlength{\tabcolsep}{3.5pt}
\begin{tabular}{lrrrr@{\hspace{10pt}}lrrrr}
\toprule
\textbf{Method} & $C^{gd}$ & $C^{pl}$ & $C^{pv}$ & $C^{to}$
& \textbf{Method} & $C^{gd}$ & $C^{pl}$ & $C^{pv}$ & $C^{to}$ \\
\midrule
\textbf{TD3}
& \textbf{220} & \textbf{2416} & \textbf{12} & \textbf{2648}
& DDPG
& 218 & 2429 & 11 & 2659 \\

RB
& 225 & 2448 & 17 & 2690
& SAC
& 218 & 2423 & 9 & 2650 \\

PPO
& 220 & 2454 & 10 & 2684
& & & & & \\
\bottomrule
\end{tabular}
\end{table}

PPO-PL also yields a 2.3\% lower $C^{svc}$ than PPO-score, suggesting that directly modelling permutations through the PL policy is better suited to the scheduling action than producing continuous scores and recovering the service order by sorting. TD3-score performs worst, possibly because sorting creates a piecewise-constant mapping from continuous scores to discrete service orders, which is poorly aligned with TD3's deterministic policy-gradient updates.

\subsection{Energy-Dispatch Policy Comparison}
\label{sec:4-2}
We next compare downstream ED controllers paired with upstream PPO-PL, the best-performing VS algorithm in Section~\ref{sec:4-1}. Four DRL algorithms and a RB controller are evaluated under the same scheduling policy. Following Section~\ref{sec:3-3}, $C^{gd}$, $C^{pl}$, and $C^{pv}$ denote the cumulative grid-purchase, pipeline (hydrogen)-purchase, and PV-curtailment costs over the evaluation horizon, respectively. We further define $C^{to}\triangleq
C^{gd}+C^{pl}+C^{pv}$. Table~\ref{tab:2} reports these costs averaged over the three test datasets and over five independently trained policies. More detailed results are provided in Appendix~\ref{app:2}. Across all controllers, hydrogen dominates the operational cost because of its higher unit price and the generally larger energy demand of HVs than EVs \cite{yao2026hierarchical}, whereas PV-curtailment cost remains small because available PV generation is preferentially used to serve EV demand (as stated in (\ref{eq:17-1})).

All three off-policy DRLs (TD3, SAC, and DDPG) outperform
the rule-based baseline, with TD3 achieving the lowest
$C^{to}$ of \pounds2,648k, followed by SAC (\pounds2,650k)
and DDPG (\pounds2,659k). In contrast, the on-policy PPO yields \pounds2,684k, only slightly below the \pounds2,690k of the rule-based controller. This supports the argument in Section~\ref{sec:3-4} that off-policy learning is better suited to the low-dimensional continuous energy-dispatch problem. It also explains why a heterogeneous, rather than homogeneous, dual-agent architecture is adopted here. Notably, the overall ordering in $C^{to}$ (TD3 $<$ SAC $<$ DDPG $<$ PPO $<$ rule-based) is consistent with that observed under heuristic scheduling in \cite{yao2026hierarchical} (Fig.~5(a) therein), despite replacing the upstream heuristic scheduler with the learned PPO-PL policy. This indicates that the relative suitability of the DRL dispatchers is broadly robust to changes in the upstream scheduling policy.

\subsection{Integrated VS-ED Performance}
\label{sec:4-3}

\begin{figure}[t]
\centering
\includegraphics[width=9.0cm, height=3.5cm,trim=1 1 1 1, clip]{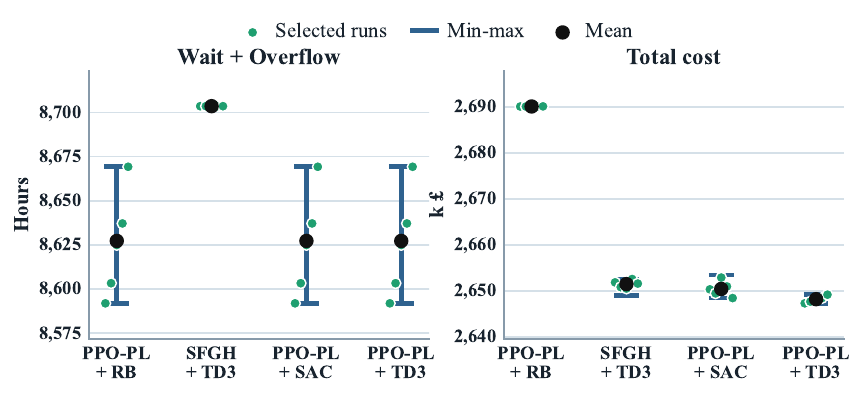}
% \figorbox{Fig1.pdf}{0.95\columnwidth}{3.2cm}
\captionsetup{font=small, skip=0pt}
\caption{Integrated performance under matched training budgets}
\label{fig:4}
\end{figure}

We then evaluate the integrated VS--ED performance under matched stage-wise training budgets. Fig.~\ref{fig:4} compares PPO-PL+RB, SFGH+TD3, PPO-PL+SAC, and PPO-PL+TD3. PPO-PL and TD3 are the best-performing VS/ED policies identified in the preceding layer-wise evaluations, while SFGH and the rule-based controller provide their respective heuristic references. To remove differences caused by independently triggered stage transitions, all four methods use the PPO-PL+TD3 schedule as the reference: Stage~I ends at epoch 5,650, Stage~II ends at epoch 12,100, and training terminates at epoch 20,000. Under this matched schedule, PPO-PL is updated during Stages~I and III, while each learned ED controller is updated during Stages~II and III; SFGH and RB remain fixed wherever used. The same five random seeds and test datasets are used for all configurations.

Under this matched schedule, PPO-PL+RB, PPO-PL+SAC, and PPO-PL+TD3 produce identical vehicle-service results. This is consistent with the directional operational structure in Section~\ref{sec:3-1}: ED determines how the scheduling-induced demand is supplied but does not alter the vehicle-service allocation or the reference VS reward. When PPO training is controlled, changing the downstream controller therefore leaves VS performance unchanged. With TD3 used for downstream dispatch, PPO-PL achieves lower vehicle-service delay than SFGH, demonstrating the benefit of learning the upstream scheduling policy. The choice of ED controller nevertheless affects energy cost: among the same PPO-PL VS, SAC and TD3 both reduce energy cost relative to RB, with TD3 achieving the lowest mean cost. These comparisons support learning the VS layer and using TD3 for downstream energy dispatch.

Table~\ref{tab:3} reports a broader comparison of 11 VS--ED methods under the plateau-triggered procedures (Section \ref{sec:3-4}), with detailed results provided in Appendix~\ref{app:2}. Unlike Fig.~\ref{fig:4}, only the maximum horizon of 20,000 epochs is common across these methods; their stage transitions are triggered independently. Different Stage-II-to-Stage-III transition epochs can therefore change the subsequent PPO update budgets and produce different final VS policies. Accordingly, the VS differences in Table~\ref{tab:3} characterise the complete training procedures rather than a direct operational effect of ED on VS. Under this protocol, PPO-PL+TD3 still achieves the lowest $C^{svc}$ and $C^{to}$ among the 11 methods, at 8,627~h and \pounds2,648k, respectively, further confirming the effectiveness of PPO-PL for VS and TD3 for ED.

The above comparison establishes the benefit of moving from the single-agent setting to the dual-agent setting. We also examine whether the directional VS$\rightarrow$ED structure should be extended to bidirectional learning. Multi-agent RL often promotes coordination through information sharing or team-level rewards \cite{lowe2017multi,wang2022individual}. We therefore test two mechanisms that introduce an ED$\rightarrow$VS learning pathway: state coupling, which adds ED information to the VS state; and reward coupling, which incorporates ED cost into the VS reward. As reported in Appendix~\ref{app:3}, both mechanisms redult in higher $C^{svc}$ and $C^{to}$ than the directional configuration.
Together with the preceding comparison against SFGH+TD3, these results show that learning both layers is beneficial, whereas the tested ED$\rightarrow$VS dependencies provide no additional benefit. This supports the adopted asymmetric dual-agent design, in which VS remains service-focused while ED adapts downstream to VS-induced demand.

\begin{table}[t]
\caption{Average vehicle-service (h) and energy-dispatch performance 
(k\pounds).}
\label{tab:3}
\centering
\footnotesize
\setlength{\tabcolsep}{2.5pt}
\renewcommand{\arraystretch}{1.0}

\begin{tabular}{@{}lrrrrrrrrrrr@{}}
\toprule
\textbf{Metric}
& \textbf{1} & \textbf{2} & \textbf{3} & \textbf{4}
& \textbf{5} & \textbf{6} & \textbf{7} & \textbf{8}
& \textbf{9} & \textbf{10} & \textbf{11} \\
\midrule

$C^{wt}$
& \textbf{7722} & 7928 & 7749 & 7749 & 7749 & 7838
& 7892 & 7936 & 7742 & 7888 & 7876 \\

$C^{ot}$
& \textbf{905} & 981 & 955 & 955 & 955 & 915
& 913 & 933 & 935 & 1174 & 950 \\

$C^{svc}$
& \textbf{8627} & 8910 & 8704 & 8704 & 8704 & 8753
& 8805 & 8870 & 8676 & 9062 & 8826 \\

\midrule

$C^{gd}$
& \textbf{220} & 223 & 245 & 223 & 225 & 225
& 220 & 218 & 218 & 218 & 218 \\

$C^{pl}$
& \textbf{2416} & 2447 & 2523 & 2447 & 2416 & 2448
& 2454 & 2429 & 2423 & 2430 & 2428 \\

$C^{pv}$
& \textbf{12} & 16 & 23 & 16 & 10 & 17
& 10 & 11 & 9 & 12 & 12 \\

$C^{to}$
& \textbf{2648 }& 2686 & 2792 & 2686 & 2651 & 2690
& 2684 & 2659 & 2650 & 2660 & 2658 \\

\bottomrule
\end{tabular}

\vspace{0.25em}

\begin{minipage}{\linewidth}
\footnotesize
\textit{Note:}
1: PPO-PL+TD3; 2: FIFO+RB; 3: SFGH+No Storage; 4: SFGH+RB;\\
5: SFGH+TD3; 6: PPO-PL+RB; 7: PPO-PL+PPO; 8: PPO-PL+DDPG;\\
9: PPO-PL+SAC; 10: TD3-score+TD3; 11: PPO-score+TD3.
\end{minipage}
\end{table}

\subsection{Three-Stage Training Ablation}
\label{sec:4-4}
We further examine the three-stage training mechanism using PPO-PL and TD3. Three variants are considered: 1) removing Stage III and allocating 10,000 epochs each to Stages I and II to provide a convergence-oriented two-stage baseline, rather than a strictly matched-budget ablation; 2) No Stage-III Reduction, which retains the Stage-II replay capacity and TD3 update intensity during Stage~III, and 3) training both policies concurrently from scratch. As shown in Table~\ref{tab:4}, the two-stage baseline without Stage III yields $C^{svc}=8,685$ h and $C^{to}=\pounds2,667$k, indicating that concurrent fine-tuning provides additional benefit beyond independently converged layer-wise training. Concurrent training from scratch further increases the service delay to 8,764~h, highlighting the benefit of first establishing competent individual policies. The No Stage-III Reduction variant increases $C^{to}$ to \pounds 2,660k. This supports the intended role of reduced replay and update intensity in mitigating the effects of scheduling-induced distribution shifts, as older ED transitions become less representative while the upstream VS policy evolves (Section~\ref{sec:3-4}). Overall, the results support the proposed three-stage mechanism for adapting ED learning to evolving VS-induced demand.

\begin{table}[t]
\caption{Ablation of the three-stage training mechanism}
\label{tab:4}
\centering
\footnotesize
\setlength{\tabcolsep}{3.2pt}
\renewcommand{\arraystretch}{0.95}
\begin{tabular}{lrrrrrrr}
\toprule
\textbf{Method}
& $C^{wt}$ & $C^{ot}$ & $C^{svc}$
& $C^{gd}$ & $C^{pl}$ & $C^{pv}$ & $C^{to}$ \\
\midrule
\textbf{PPO-PL+TD3}
& \textbf{7722} & \textbf{905} & \textbf{8627}
& \textbf{220} & \textbf{2416} & \textbf{12} & \textbf{2648} \\
No Stage III
& 7744 & 942 & 8685 & 220 & 2435 & 12 & 2667 \\
No Stage-III Reduction
& 7722 & 905 & 8627 & 220 & 2427 & 12 & 2660 \\
Concurrent Scratch
& 7837 & 927 & 8764 & 221 & 2435 & 12 & 2667 \\
\bottomrule

\multicolumn{8}{@{}l@{}}{%
  \textit{Note:} h for VS and k\pounds\ for ED.
}\\
\end{tabular}
\end{table}

\subsection{Operating Analysis}\label{sec:4-5}

\begin{figure}[t]
\centering
\includegraphics[width=8.5cm, trim=1 1 1 1, clip]{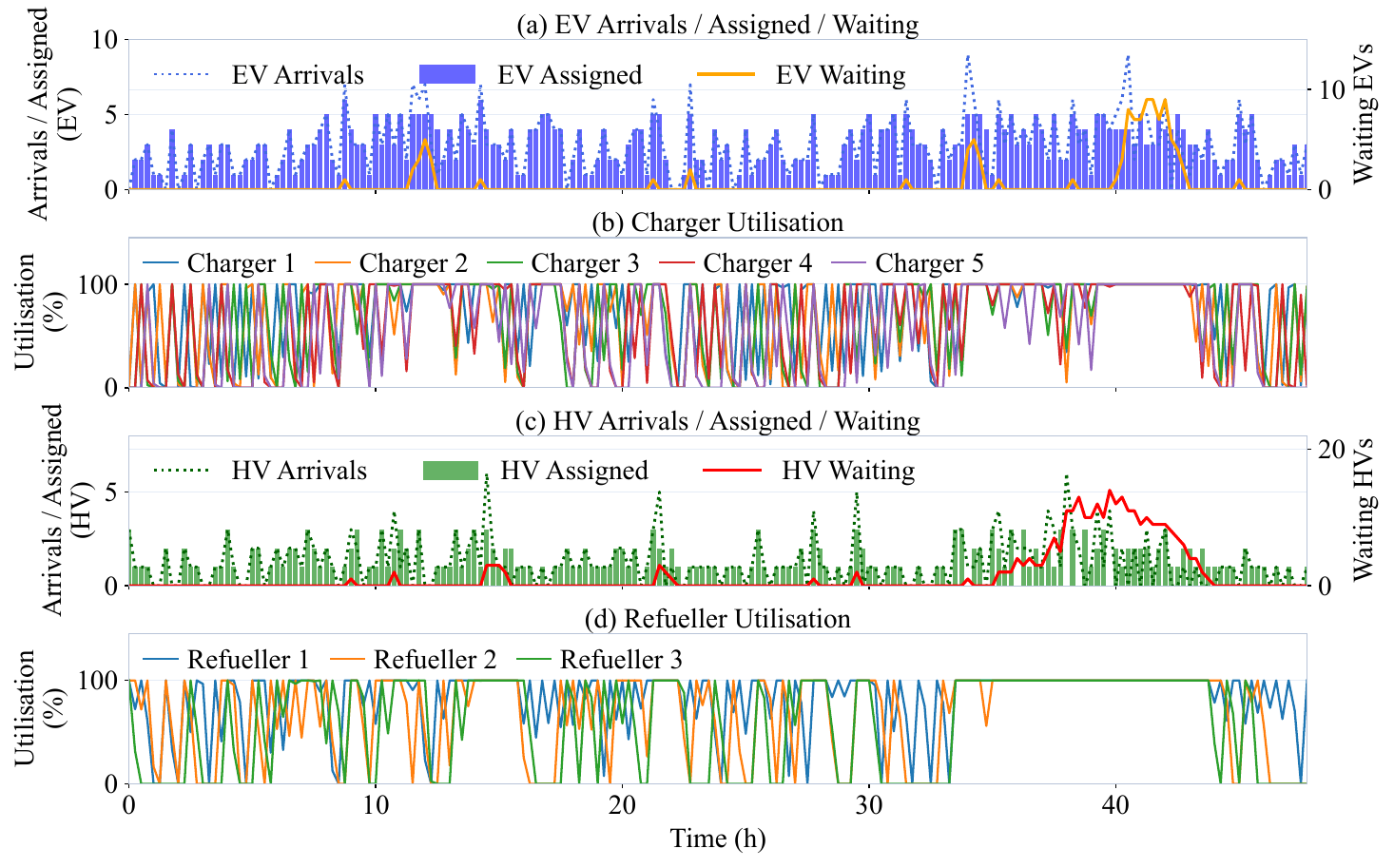}
% \figorbox{Fig1.pdf}{0.95\columnwidth}{3.2cm}
\captionsetup{font=small,skip=0pt}
\caption{Two-day vehicle-service operation and pile utilisation}
\label{fig:operating-VS}
\end{figure}
\begin{figure}[t]
\centering
\includegraphics[width=8.5cm, trim=1 1 1 1, clip]{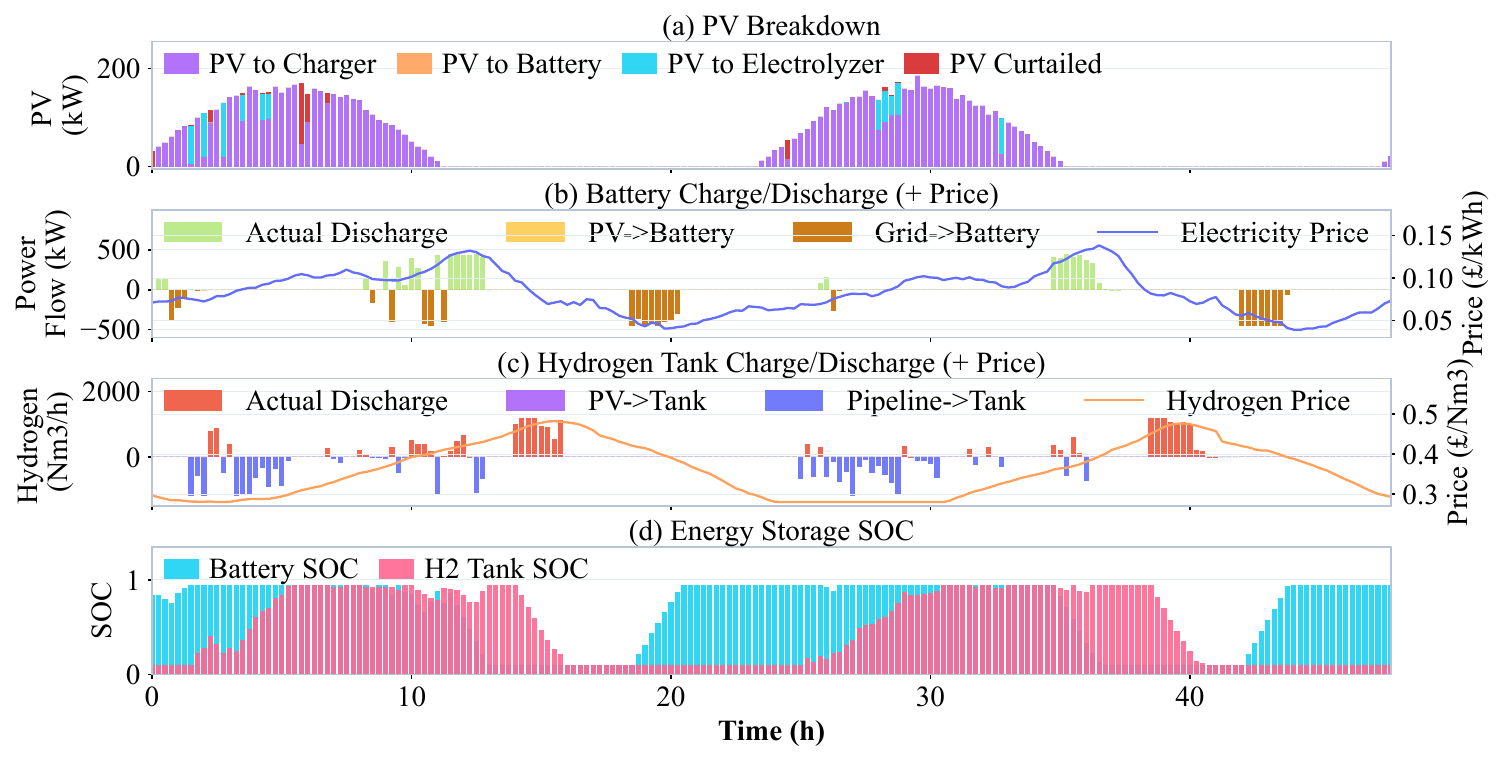}
% \figorbox{Fig1.pdf}{0.95\columnwidth}{3.2cm}
\captionsetup{font=small,skip=0pt}
\caption{Two-day renewable breakdown and storage operation}
\label{fig:operating-ED}
\end{figure}
\begin{figure}[b]
\centering
\includegraphics[width=8.5cm, trim=1 1 1 1, clip]{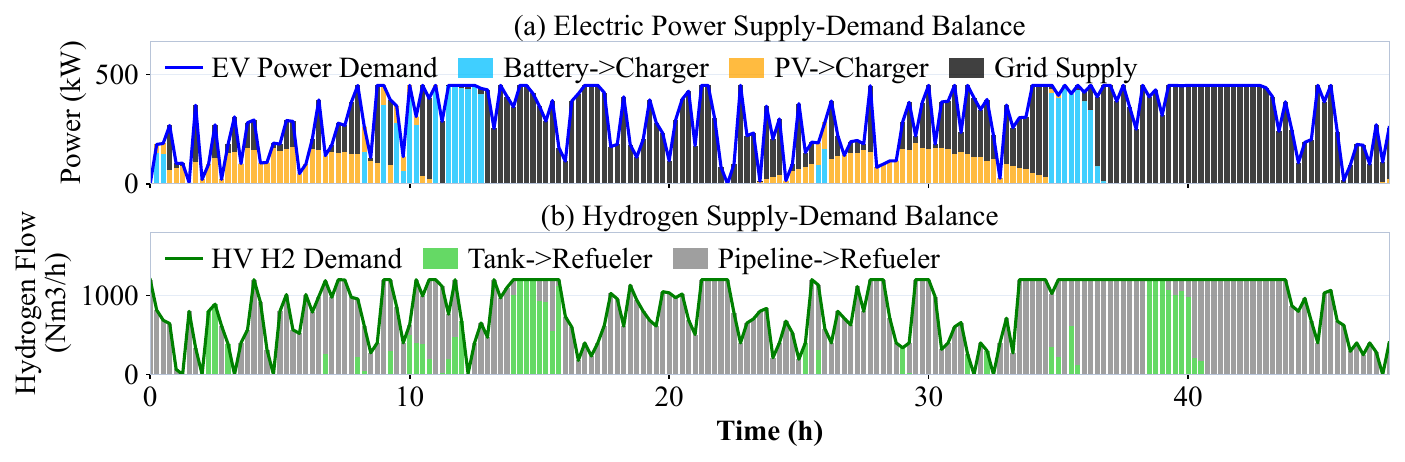}
% \figorbox{Fig1.pdf}{0.95\columnwidth}{3.2cm}
\captionsetup{font=small,skip=0pt}
\caption{Two-day supply--demand balance}
\label{fig:operating-balance}
\end{figure}

To illustrate the operating behaviour of the proposed framework, Figs.~\ref{fig:operating-VS}--\ref{fig:operating-balance} show a representative two-day trajectory covering vehicle scheduling, energy dispatch, and the resulting supply--demand balance. The corresponding year-long operating trajectories can be explored through the accompanying interactive dashboard~\cite{Dual-agent-RL-EHTS-Dashboard}. Rather than comparing aggregate performance above, this analysis examines how the learned policies respond temporally to variations in vehicle arrivals, service capacity, renewable generation, and energy prices.

On the VS side, Fig.~\ref{fig:operating-VS} shows that waiting is concentrated around periods of elevated vehicle arrivals and high charger/refueller utilisation, rather than accumulating continuously. In particular, the pronounced demand around 40--43 h produces temporary EV and HV queues, which are subsequently cleared as the load subsides. The frequent saturation of individual chargers and refuellers during such periods further indicates that the observed queues are associated with temporary service-capacity pressure. Overall, the learned VS policy dynamically allocates the available service resources under time-varying demand, with temporary queues being cleared as the service pressure subsides.

On the energy side, Fig.~\ref{fig:operating-ED} demonstrates clear temporal coordination between renewable generation, storage, and external energy procurement. Available PV is predominantly supplied directly to EV charging, with only limited curtailment, while the battery charges during low-price periods and discharges when electricity prices are higher. A similar pattern is observed for hydrogen: the tank is replenished from the pipeline during lower-price periods and discharged during higher-price periods. The corresponding SOC trajectories confirm that both storage devices are actively cycled to shift energy across time rather than being operated myopically. Finally, Fig.~\ref{fig:operating-balance} shows that EV demand is jointly supplied by PV, battery discharge, and the grid, whereas HV demand is met by tank discharge and pipeline supply. The external grid and hydrogen pipeline therefore absorb the residual demand when local renewable or stored energy is insufficient, maintaining supply--demand balance throughout the two-day trajectory. Together, these results illustrate how the proposed framework combines dynamic vehicle scheduling with price-aware energy management while respecting the physical coupling between service demand and downstream energy supply.
\subsection{Generalisation under Unseen Operating Conditions}
\label{sec:4-6}

\begin{figure}[t]
\centering
\includegraphics[width=9cm, trim=1 1 1 1, clip]{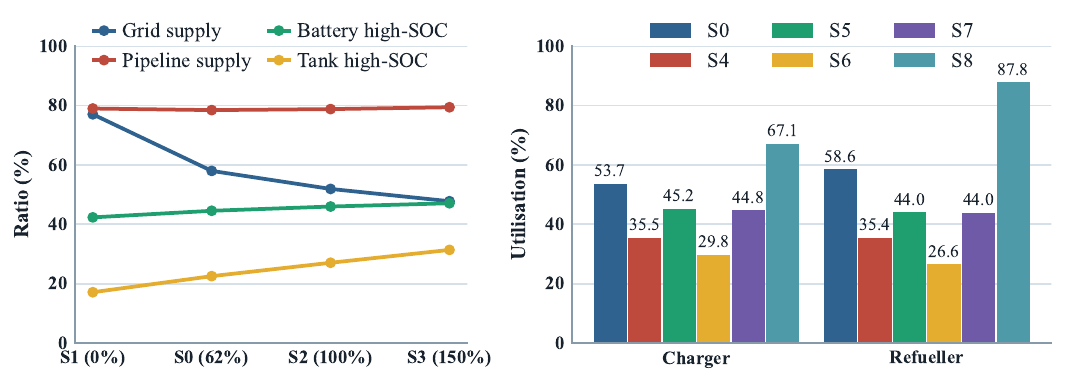}
% \figorbox{Fig1.pdf}{0.95\columnwidth}{3.2cm}
\captionsetup{font=small,skip=0pt}
\caption{Operating characteristics under unseen conditions}
\label{fig:8}
\end{figure}

The proposed framework is further evaluated under eight unseen operating conditions, denoted S1--S8 (as shown in Table \ref{tab:5}), against the nominal condition S0. Interactive comparisons across all eight unseen conditions are also available through the accompanying dashboard~\cite{Dual-agent-RL-EHTS-Dashboard}. S0 uses 62\% PV penetration (the ratio of PV capacity to peak demand), the high-demand setting, and five EV chargers/three HV refuellers. S1--S3 vary the PV penetration to 0\%, 100\%, and 150\%, respectively; S4--S6 \cite{yao2026hierarchical} reduce vehicle arrivals (medium-demand), per-vehicle demand (medium-demand), and both simultaneously (low-demand); and S7--S8 increase and reduce the service infrastructure to 6/4  to 4/2 chargers/refuellers, respectively. The same fixed-dimensional VS-state representation is retained in these scenarios, with unavailable pile slots zero-padded where required. All scenarios directly deploy the trained PPO-PL and TD3 policies without retraining or fine-tuning. Results are averaged over five independent runs, with S0--S3 and S7--S8 further averaged over the three test datasets. Additional comparisons with the fully heuristic FIFO+RB baseline and the single-agent learning SFGH+TD3 framework are provided in Appendix~\ref{app:4}, where PPO-PL+TD3 shows stronger overall generalisation across the unseen conditions. Fig.~\ref{fig:8} shows the corresponding operating characteristics under unseen conditions, including the grid-supply ratio (relative to demand), pipeline-supply ratio, high-SOC occupancy (SOC $\geq 0.90$) of the battery and tank, and charger and refueller utilisation.

\textbf{1) PV penetration variation:} The invariant VS performance (see Table \ref{tab:5}) across S0--S3 is expected from the adopted one-way coupling and confirms implementation consistency. More importantly, the ED policy adapts to the unseen PV conditions without retraining: as PV penetration increases from 0\% (S1) to 150\% (S3), the grid-supply ratio decreases from 77.0\% to 47.8\%, while both storage devices exhibit increased high-SOC occupancy (see Fig.~\ref{fig:8}). The nearly unchanged pipeline ratio suggests that the ED policy accommodates PV variations mainly through adjustments on the electricity and storage sides. Interestingly, at higher penetration levels, increased PV curtailment offsets part of the benefit of additional PV generation, explaining why further PV deployment does not continuously reduce $C^{ to}$. Overall, these results demonstrate that the trained ED policy can adapt its energy-management behaviour across the tested PV conditions without retraining.

\textbf{2) Transport-demand variation:}
Reducing vehicle arrivals (S4) decreases $C^{svc}$ to 628~h, whereas reducing per-vehicle demand (S5) results in 2,240~h; the corresponding $C^{to}$ decrease to \pounds1,582k and \pounds1,996k, respectively. When both are reduced simultaneously (S6), $C^{svc}$ and $C^{to}$ further decrease to 239~h and \pounds1,187k (see Table \ref{tab:5}). The pile-utilisation results in Fig.~\ref{fig:8} provide further insight: EV/HV utilisation decreases to 35.5\%/35.4\% when vehicle arrivals are reduced, compared with 45.2\%/44.0\% when only per-vehicle demand is reduced. These results suggest that service congestion is more sensitive to vehicle-arrival intensity under the studied conditions, while the trained policies remain effective across all three unseen lower-demand scenarios without retraining.

\textbf{3) Service-infrastructure variation:}
Increasing the infrastructure from the nominal $5/3$ to $6/4$ chargers/refuellers (S7) reduces $C^{svc}$ from 8,627 to 1,796~h, while $C^{to}$ remains similar (see Table \ref{tab:5}). Meanwhile, Fig.~\ref{fig:8} shows that EV/HV pile utilisation decreases from 53.7\%/58.6\% to 44.8\%/44.0\%, indicating that the trained priority policy can accommodate the additional service capacity without retraining. In contrast, reducing the infrastructure to $4/2$ (S8) sharply increases $C^{svc}$ to 209,163~h, with HV-refueller utilisation reaching 87.8\%, revealing a severe service-capacity bottleneck. Thus, the policy remains deployable under the reduced configuration, but does not preserve performance under severe capacity insufficiency.

\begin{table}[t]
\centering
\caption{Generalisation performance under S0--S8.}
\label{tab:5}
\footnotesize
\setlength{\tabcolsep}{2pt}
\renewcommand{\arraystretch}{1.0}
\begin{tabular}{@{}clrrrrrrr@{}}
\toprule
\textbf{Scenario} & \textbf{Setting}
& $C^{wt}$ & $C^{ot}$ & $C^{svc}$
& $C^{gd}$ & $C^{pl}$ & $C^{pv}$ & $C^{to}$ \\
\midrule
\textbf{S0} & \textbf{Nominal}      & \textbf{7722 }  & \textbf{905 }  & \textbf{8627}   & \textbf{220} & \textbf{2416} & \textbf{12}& \textbf{2648} \\
S1 & PV: 0\%      & 7722   & 905   & 8627   & 296 & 2443 & 0  & 2739 \\
S2 & PV: 100\%    & 7722   & 905   & 8627   & 195 & 2418 & 37 & 2649 \\
S3 & PV: 150\%    & 7722   & 905   & 8627   & 177 & 2411 & 89 & 2677 \\
S4 & Arrivals $\downarrow$    & 626    & 2     & 628    & 131 & 1428 & 23 & 1582 \\
S5 & Demand  $\downarrow$     & 2211   & 29    & 2240   & 179 & 1802 & 16 & 1996 \\
S6 & Both  $\downarrow$       & 239    & 0     & 239    & 106 & 1052 & 29 & 1187 \\
S7 & Infra.: 6/4  & 1778   & 19    & 1796   & 223 & 2428 & 12 & 2663 \\
S8 & Infra.: 4/2  & 116647 & 92517 & 209163 & 216 & 2378 & 11 & 2605 \\
\bottomrule
\multicolumn{8}{@{}l@{}}{%
  \textit{Note:} h for VS and k\pounds\ for ED.
}\\
\end{tabular}
\end{table}

%=============================================================================
\section{Conclusion}
\label{sec:5}
%=============================================================================
This paper presented a sequential dual-agent framework for vehicle scheduling (VS) and energy dispatch (ED) in an integrated electric--hydrogen--transport system. The framework combines PPO with a Plackett--Luce policy (PPO--PL) for the factorial scheduling action space and TD3 for continuous energy dispatch. A three-stage training mechanism was developed to address scheduling-induced changes in energy demand through layer-wise pretraining and concurrent policy updates. Year-long experiments showed that PPO--PL+TD3 achieved the lowest vehicle-service delay and energy cost among many other evaluated methods, demonstrating the benefits of jointly learning VS and ED. The trained policies also maintained their performance under unseen PV penetration, transport demand, and service-infrastructure scenarios. Future work will incorporate dynamic equipment degradation, life-cycle carbon emissions, and bidirectional grid exchange to support more sustainable EHTS operation.

\bibliographystyle{IEEEtran}
\bibliography{ehts-refs}

%===============================================================
\appendices

\section{Three-Stage Training Procedure}
\label{app:1}

Let $E_{\max}$ denote the maximum number of training epochs,
$T_{roll}$ denote the number of control intervals in each rollout,
$q\in\{1,2,3\}$ denote the current training stage, and $n_{pl}$
denote the number of consecutive plateau checks.

\begin{algorithm}[t]
\caption{Three-stage training of the dual-agent RL framework}
\label{alg:train}
\begin{algorithmic}[1]

\State Initialise PPO $(\pi_\theta,V_\psi)$, TD3
$(\mu_\phi,Q_{\theta_1},Q_{\theta_2})$, and the ED replay buffer
\State Set the training stage $q\leftarrow1$

\For{$e=1,\ldots,E_{\max}$}

    \For{$t=1,\ldots,T_{roll}$}

        \State Construct the VS state $\mathbf{s}^{vs}_t$
        according to \eqref{eq:23}--\eqref{eq:25}

        \If{$q=2$}
            \State Generate $\mathbf{a}^{vs}_t$ greedily using the
            frozen PPO policy
        \Else
            \State Sample $\mathbf{a}^{vs}_t$ using the PL policy
            in \eqref{eq:21} and the action definition in \eqref{eq:26}
        \EndIf

        \State Execute $\mathbf{a}^{vs}_t$ and obtain
        $(P_t^{ev},H_t^{hv})$ according to
        \eqref{eq:7}

        \State Construct the ED state $\mathbf{s}^{ed}_t$
        according to \eqref{eq:30}--\eqref{eq:31}

        \If{$q=1$}
            \State Determine $\mathbf{a}^{ed}_t$ using the fixed
            rule-based ED controller in \cite{yao2026hierarchical}
        \Else
            \State Generate $\mathbf{a}^{ed}_t$ using the TD3 actor and map it
to feasible dispatch decisions according to
\eqref{eq:32}--\eqref{eq:35}
        \EndIf

        \State Execute $\mathbf{a}^{ed}_t$ and obtain
        $r_t^{vs}$, $r_t^{ed}$, $\mathbf{s}^{vs}_{t+1}$, and
        $\mathbf{s}^{ed}_{t+1}$

        \If{$q\in\{1,3\}$}
            \State Store the VS transition for PPO updating
        \EndIf

        \If{$q\in\{2,3\}$}
            \State Store the ED transition in the replay buffer
        \EndIf

    \EndFor

    \If{$q\in\{1,3\}$}
        \State Update PPO according to
        \cite{schulman2017proximal} %\eqref{eq:28}--\eqref{eq:29}
    \EndIf

    \If{$q\in\{2,3\}$}
        \State Update TD3 according to
        \cite{fujimoto2018addressing} %\eqref{eq:38}--\eqref{eq:39}
    \EndIf

    \If{$q<3$ and the plateau criterion is satisfied}
        \State $q\leftarrow q+1$
        \If{$q=3$}
            \State Reduce the ED replay capacity to $10\%$ of its Stage~II value
            \State Reduce the TD3 updates per rollout to one-third of its Stage~II value
        \EndIf
    \EndIf
\EndFor

\end{algorithmic}
\end{algorithm}

\section{Experimental Setup}
\label{app:1-1}

\textbf{Dataset Construction:}
The experiments use hybrid annual datasets combining real-world meteorological and electricity-price data with synthetically generated hydrogen-price and transport-demand profiles. All datasets are represented at a 15-min resolution ($\Delta t=0.25$~h), with each annual dataset containing 35,040 decision intervals. Four high-demand datasets are constructed, with one used for training and three independently generated datasets (Test~1--3) retained exclusively for testing. The meteorological inputs consist of solar irradiance and ambient temperature for South Wales, while the electricity-price profile is based on the GB Market Index Price. 

Following \cite{yao2026hierarchical}, PV generation is calculated from solar irradiance and ambient temperature using the temperature-dependent PV model; hydrogen prices are generated using a stochastic temporal model incorporating seasonal, weekly, intra-day, electricity-price-coupled, and stochastic variations; EV and HV arrivals are generated using non-homogeneous Poisson processes with time-varying arrival intensities; and individual charging and refuelling requirements are generated from stochastic demand models. Specifically, under the high-demand setting, the stochastic data-generation models use EV and HV baseline arrival-scale parameters of 5 and 4, respectively, and baseline individual service-demand parameters of 25~kWh and 180~Nm$^3$, respectively. These values are model inputs rather than fixed arrival counts or realised individual service demands. The medium-arrival condition reduces the EV and HV arrival-scale parameters to 4 and 3, respectively, while retaining the high-demand service-demand parameters. The medium-demand condition reduces the EV and HV baseline service-demand parameters to 20~kWh and 150~Nm$^3$, respectively, while retaining the high-demand arrival-scale parameters. The low-demand condition combines both parameter reductions.

\textbf{Physical Configuration:}
The nominal EHTS contains $N^{ec}=5$ EV chargers and $N^{hr}=3$ hydrogen refuellers. The effective pile-to-vehicle service efficiency is $\eta^{c}=0.99$ for both vehicle classes. The battery capacity and power limit are $E^{bt,cp}=900$~kWh and $P^{bt,max}=450$~kW, respectively. Its charging and discharging efficiencies are $\eta^{bt,ch}=\eta^{bt,dc}=0.95$, and its SOC limits are $SOC^{bt,min}=0.10$ and $SOC^{bt,max}=0.95$. The hydrogen-tank capacity is $H^{tk,cp}=2400$~Nm$^3$. Its charging and discharging efficiencies are $\eta^{tk,ch}=\eta^{tk,dc}=0.995$, and its SOC limits are $SOC^{tk,min}=0.10$ and $SOC^{tk,max}=0.95$. The maximum tank charging/discharging rate is 1200~Nm$^3$/h, corresponding to a maximum amount of $H^{tk,max}=300$~Nm$^3$ per decision interval at $\Delta t=0.25$~h. The electrolyser electricity consumption is $\eta^{el}=5.0$~kWh/Nm$^3$. These physical parameters remain fixed unless explicitly varied in the generalisation experiments.

The prescribed waiting-time thresholds are $w^{ev,\max}=0.50$~h and $w^{hv,\max}=0.75$~h. At each decision interval, at most $M=10$ EVs and $N=6$ HVs are retained as scheduling candidates. Unused candidate positions are zero-padded and masked from the PL permutation. The pile-state representation retains a fixed input dimension across the service-infrastructure generalisation experiments, with unavailable charger/refueller positions zero-padded and excluded from physical allocation.

\textbf{DRL Settings:}
The PPO-PL actor and critic, together with the TD3 actor and both critics, each use two fully connected hidden layers with 256 neurons per layer. Both agents use a discount factor of 0.995. For PPO-PL, the GAE parameter is 0.95 and the clipping ratio is 0.15. Four update iterations are performed per epoch with a mini-batch size of 128. The actor and critic learning rates are $5\times10^{-5}$ and $10^{-4}$, respectively, and the initial entropy coefficient is 0.004. The VS reward weights are $\omega^{wt}=\omega^{ot}=1$. For TD3, the actor and critic learning rates are both $3\times10^{-4}$, the soft target-update coefficient is 0.005, and the mini-batch size is 256. Target-policy smoothing uses Gaussian noise with a standard deviation of 0.15, clipped to $[-0.3,0.3]$. Uniformly random actions are used during the first 2000 TD3 interaction steps, after which Gaussian exploration noise with a standard deviation of 0.15 is added to the actor output. The replay-buffer capacity is 500,000 transitions in Stage II and 50,000 in Stage III. During Stage III, the TD3 update intensity is reduced to one-third of its Stage-II value.

The ED state uses $L=6$ historical observations with a stride of four decision intervals, covering the preceding six hours. All continuous VS and ED state variables are normalised to $[0,1]$ using fixed min--max scaling, with the same bounds retained during training and all testing scenarios. All DRL configurations are trained using five independent random seeds, 42--46. The experiments are implemented in PyTorch, with independent runs distributed across 14 homogeneous laboratory PCs, each equipped with an Intel Core i5-10500 CPU (3.10~GHz) and 16~GB RAM.

\section{Additional Experimental Results}
\label{app:2}
%=============================================================================

Table~\ref{tab:app-1} reports dataset-specific results for the baseline and ablation comparisons in Section~\ref{sec:4} and Appendix~\ref{app:3} under their respective three-stage training procedures. The matched-budget results in Fig.~\ref{fig:4} are not included here. The DRL policies are evaluated over five independently trained runs using random seeds 42--46, while the deterministic baselines are evaluated once. For each test set, the VS-side columns report $C^{wt}$, $C^{ot}$, and $C^{svc}$ in h, while the ED-side columns report $C^{gd}$, $C^{pl}$, $C^{pv}$, and $C^{to}$ in k\pounds. Avg denotes the corresponding average over Test~1, Test~2, and Test~3 \cite{yao2026hierarchical}.

\begin{table*}[t]
\caption{Complete VS and ED performance over the three held-out test sets}
\label{tab:app-1}
\centering
\begingroup
\tiny
\setlength{\tabcolsep}{1.25pt}
\renewcommand{\arraystretch}{0.78}
\resizebox{\textwidth}{!}{%
\begin{tabular}{@{}lrrrrrrrrrrrrrr@{}}
\toprule
\multirow{3}{*}{\textbf{Method}} & \multicolumn{7}{c}{\textbf{Test 1}} & \multicolumn{7}{c}{\textbf{Test 2}} \\
\cmidrule(lr){2-8}\cmidrule(lr){9-15}
 & \multicolumn{3}{c}{\textbf{VS side (h)}} & \multicolumn{4}{c}{\textbf{ED side (k\pounds)}} & \multicolumn{3}{c}{\textbf{VS side (h)}} & \multicolumn{4}{c}{\textbf{ED side (k\pounds)}} \\
\cmidrule(lr){2-4}\cmidrule(lr){5-8}\cmidrule(lr){9-11}\cmidrule(lr){12-15}
 & $C^{wt}$ & $C^{ot}$ & $C^{svc}$ & $C^{gd}$ & $C^{pl}$  & $C^{pv}$ & $C^{to}$ & $C^{wt}$ & $C^{ot}$ & $C^{svc}$ & $C^{gd}$ & $C^{pl}$ & $C^{pv}$ & $C^{to}$ \\
\midrule
PPO-PL+TD3 & 7738 & 981 & 8719 & 222 & 2405 & 12 & 2640 & 7906 & 893 & 8799 & 218 & 2448 & 12 & 2678 \\
FIFO+RB & 7949 & 1056 & 9005 & 225 & 2436 & 16 & 2676 & 8112 & 971 & 9083 & 221 & 2480 & 16 & 2717 \\
SFGH+No Storage & 7762 & 1028 & 8790 & 247 & 2512 & 24 & 2782 & 7937 & 944 & 8881 & 243 & 2556 & 24 & 2822 \\
SFGH+RB & 7762 & 1028 & 8790 & 225 & 2436 & 16 & 2676 & 7937 & 944 & 8881 & 221 & 2480 & 16 & 2717 \\
SFGH+TD3 & 7762 & 1028 & 8790 & 227 & 2404 & 11 & 2642& 7937 & 944 & 8881 & 224 & 2448 & 10 & 2682 \\
PPO-PL+RB & 7857 & 996 & 8853 & 227 & 2437 & 17 & 2680 & 8023 & 898 & 8922 & 223 & 2481 & 17 & 2721 \\
No Stage III & 7748 & 1005 & 8754 & 222 & 2424 & 12 & 2658 & 7942 & 942 & 8884 & 219 & 2468 & 12 & 2699 \\
No Stage-III Reduction & 7738 & 981 & 8719 & 222 & 2416 & 13 & 2651 & 7906 & 893 & 8799 & 219 & 2459 & 12 & 2690 \\
Concurrent Scratch & 7852 & 1000 & 8852 & 223 & 2425 & 12 & 2659 & 8026 & 917 & 8943 & 219 & 2466 & 12 & 2697 \\
PPO-PL+PPO & 7908 & 987 & 8896 & 221 & 2443 & 10 & 2674 & 8081 & 900 & 8981 & 218 & 2486 & 10 & 2714 \\
PPO-PL+DDPG & 7960 & 1021 & 8981 & 220 & 2418 & 12 & 2649 & 8125 & 916 & 9041 & 216 & 2461 & 12 & 2689 \\
PPO-PL+SAC & 7751 & 1002 & 8753 & 220 & 2412 & 10 & 2641 & 7930 & 926 & 8856 & 216 & 2455 & 10 & 2681 \\
TD3-score +TD3 & 7899 & 1250 & 9149 & 220 & 2418 & 12 & 2651 & 8075 & 1170 & 9245 & 216 & 2461 & 12 & 2689 \\
PPO-score VS+TD3 & 7896 & 1023 & 8919 & 220 & 2416 & 12 & 2648 & 8055 & 937 & 8992 & 216 & 2460 & 12 & 2688 \\
Reward Coupling ($\alpha=0.10$) & 7843 & 989 & 8832 & 219 & 2418 & 12 & 2649 & 8006 & 895 & 8901 & 216 & 2461 & 12 & 2688 \\
State Coupling ($\alpha=0$) & 7789 & 996 & 8785 & 221 & 2418 & 10 & 2650 & 7974 & 921 & 8896 & 217 & 2461 & 10 & 2688 \\
State Coupling ($\alpha=0.1$) & 7819 & 1003 & 8821 & 221 & 2420 & 11 & 2652 & 7989 & 915 & 8905 & 217 & 2462 & 10 & 2690 \\
Reward Coupling ($\alpha=0.20$) & 7815 & 1022 & 8837 & 221 & 2415 & 12 & 2648 & 7982 & 935 & 8917 & 217 & 2458 & 12 & 2688 \\
State Coupling ($\alpha=0.2$) & 7862 & 1026 & 8888 & 221 & 2416 & 11 & 2649 & 8033 & 941 & 8974 & 217 & 2459 & 11 & 2687 \\

\bottomrule
\end{tabular}%
}
\vspace{0.25em}
\resizebox{\textwidth}{!}{%
\begin{tabular}{@{}lrrrrrrrrrrrrrr@{}}
\toprule
\multirow{3}{*}{\textbf{Method}} & \multicolumn{7}{c}{\textbf{Test 3}} & \multicolumn{7}{c}{\textbf{Avg}} \\
\cmidrule(lr){2-8}\cmidrule(lr){9-15}
 & \multicolumn{3}{c}{\textbf{VS side (h)}} & \multicolumn{4}{c}{\textbf{ED side (k\pounds)}} & \multicolumn{3}{c}{\textbf{VS side (h)}} & \multicolumn{4}{c}{\textbf{ED side (k\pounds)}} \\
\cmidrule(lr){2-4}\cmidrule(lr){5-8}\cmidrule(lr){9-11}\cmidrule(lr){12-15}
&$C^{wt}$ & $C^{ot}$ & $C^{svc}$ & $C^{gd}$ & $C^{pl}$  & $C^{pv}$ & $C^{to}$ & $C^{wt}$ & $C^{ot}$ & $C^{svc}$ & $C^{gd}$ & $C^{pl}$ & $C^{pv}$ & $C^{to}$ \\
\midrule
PPO-PL+TD3 & 7521 & 843 & 8364 & 220 & 2395 & 12 & 2626 & 7722 & 905 & 8627 & 220 & 2416 & 12 & 2648 \\
FIFO+RB & 7724 & 918 & 8642 & 223 & 2427 & 16 & 2666 & 7928 & 981 & 8910 & 223 & 2447 & 16 & 2686 \\
SFGH+No Storage & 7548 & 892 & 8440 & 245 & 2502 & 23 & 2770 & 7749 & 955 & 8704 & 245 & 2523 & 23 & 2792 \\
SFGH+RB & 7548 & 892 & 8440 & 223 & 2427 & 16 & 2666 & 7749 & 955 & 8704 & 223 & 2447 & 16 & 2686 \\
SFGH+TD3 & 7548 & 892 & 8440 & 224 & 2396 & 10 & 2630 & 7749 & 955 & 8704 & 225 & 2416 & 10 & 2651 \\
PPO-PL+RB & 7634 & 850 & 8483 & 225 & 2428 & 17 & 2669 & 7838 & 915 & 8753 & 225 & 2448 & 17 & 2690 \\
No Stage III & 7540 & 878 & 8418 & 220 & 2414 & 11 & 2645 & 7744 & 942 & 8685 & 220 & 2435 & 12 & 2667 \\
No Stage-III Reduction & 7521 & 843 & 8364 & 220 & 2407 & 12 & 2639 & 7722 & 905 & 8627 & 220 & 2427 & 12 & 2660 \\
Concurrent Scratch & 7633 & 864 & 8497 & 221 & 2414 & 11 & 2646 & 7837 & 927 & 8764 & 221 & 2435 & 12 & 2667 \\
PPO-PL+PPO & 7686 & 852 & 8538 & 220 & 2434 & 10 & 2663 & 7892 & 913 & 8805 & 220 & 2454 & 10 & 2684 \\
PPO-PL+DDPG & 7723 & 863 & 8587 & 217 & 2410 & 11 & 2638 & 7936 & 933 & 8870 & 218 & 2429 & 11 & 2659 \\
PPO-PL+SAC & 7545 & 876 & 8421 & 217 & 2403 & 9 & 2629 & 7742 & 935 & 8676 & 218 & 2423 & 9 & 2650 \\
TD3-score +TD3 & 7690 & 1103 & 8794 & 217 & 2410 & 12 & 2639 & 7888 & 1174 & 9062 & 218 & 2430 & 12 & 2660 \\
PPO-score +TD3 & 7677 & 890 & 8567 & 217 & 2407 & 12 & 2636 & 7876 & 950 & 8826 & 218 & 2428 & 12 & 2658 \\
Reward Coupling ($\alpha=0.10$) & 7618 & 847 & 8464 & 217 & 2409 & 11 & 2637 & 7822 & 910 & 8733 & 217 & 2429 & 12 & 2658 \\
State Coupling ($\alpha=0$) & 7577 & 864 & 8441 & 218 & 2409 & 10 & 2637 & 7780 & 927 & 8707 & 219 & 2429 & 10 & 2658 \\
State Coupling ($\alpha=0.1$) & 7599 & 864 & 8463 & 219 & 2410 & 10 & 2640 & 7802 & 927 & 8730 & 219 & 2431 & 10 & 2661 \\
Reward Coupling ($\alpha=0.20$) & 7592 & 884 & 8476 & 219 & 2405 & 11 & 2635 & 7797 & 947 & 8743 & 219 & 2426 & 12 & 2657 \\
State Coupling ($\alpha=0.2$) & 7644 & 890 & 8534 & 219 & 2406 & 10 & 2635 & 7846 & 952 & 8798 & 219 & 2427 & 11 & 2657 \\

\bottomrule
\end{tabular}%
}
\endgroup
\end{table*}

\section{Coupling Mechanism Ablation}
\label{app:3}
We examine two optional mechanisms that introduce an additional ED$\rightarrow$VS learning pathway into the proposed dual-agent framework. The first pathway is state coupling, where the VS agent is allowed to observe energy-side states when making scheduling decisions. The second pathway is reward coupling, where the VS reward is augmented with an energy-side cost signal so that the scheduler is encouraged to generate vehicle-service decisions that are not only locally efficient for queues, but also easier for the downstream energy dispatcher to support. This subsection examines whether additional state or reward coupling benefits the proposed dual-agent framework.

Let $\textbf{o}_t^{vs}$ denote the native vehicle-service state observed by the scheduler, including queue, service, and candidate-assignment information, and let $\textbf{o}_t^{ed}$ denote the energy-side state observed by the dispatcher, such as storage states, renewable generation, load-related information, and energy-price conditions. In the isolated setting (proposed in Section \ref{sec:3-2}), the VS state is restricted to the vehicle-service observation only, i.e., $\textbf{s}_t^{vs} = \textbf{o}_t^{vs}$.
With state coupling enabled, the scheduler additionally receives the energy-side observation, i.e., $\textbf{s}_t^{vs} = \left[\textbf{o}_t^{vs}, \textbf{o}_t^{ed}\right]$. 
This design allows the VS agent to condition its scheduling decisions on the downstream energy context, rather than treating the ED layer as an unobserved environment response.

Reward coupling is controlled by a scalar coefficient $\alpha$. Without reward coupling (proposed in Section \ref{sec:3-2}), the VS reward only reflects vehicle-service performance, i.e., $r_t^{vs} = r_t^{vs,svc}$. When reward coupling is enabled, the VS reward includes an additional normalised energy-side cost term:
\begin{equation}
r_t^{vs} =
r_t^{v,svc}
-
\alpha \, \widehat{C}_{t}^{ed},
\end{equation}
where $\widehat{C}_{t}^{ed}$ denotes the normalised ED-side operational-cost signal. The coefficient $\alpha$ therefore controls how strongly the scheduler is penalized for decisions that increase downstream energy cost. Table~\ref{tab:coupling-results} compares the proposed setting (Variant 1) with five coupling variants, where $\alpha=0$ corresponds to no reward coupling, while $\alpha=0.10$ and $\alpha=0.20$ represent two progressively stronger reward-coupling strengths.

The proposed reference setting achieves the lowest $C^{wt}$, $C^{ot}$, $C^{svc}$, and $C^{to}$ among the evaluated variants. All five coupling variants increase both overall service delay and energy cost relative to the reference. Increasing $\alpha$ from 0.10 to 0.20 slightly reduces the reported $C^{to}$ in both settings but increases $C^{svc}$. One possible explanation is that feeding the ED-side objective back into the VS reward introduces an additional learning dependency from ED to VS, potentially turning the original one-way non-stationarity into a more complex bidirectional interaction during  training. The coupling variants yield energy costs ranging from \pounds2,657k to \pounds2,661k, compared with \pounds2,648k for the reference. Overall, the tested state and reward coupling mechanisms worsen both VS and ED performance under the evaluated conditions, supporting the adopted asymmetric dual-agent design.

\begin{table}[t]
\caption{Coupling-mechanism ablation of the proposed framework
(h for VS and k\pounds\ for ED).}
\label{tab:coupling-results}
\centering
\footnotesize
\setlength{\tabcolsep}{2.5pt}
\renewcommand{\arraystretch}{1.0}

\begin{tabular}{@{}cccrrrrrrr@{}}
\toprule
\textbf{Variant} & \textbf{Reward ($\alpha$)} & \textbf{State}
& $C^{wt}$ & $C^{ot}$ & $C^{svc}$
& $C^{gd}$ & $C^{pl}$ & $C^{pv}$ & $C^{to}$ \\
\midrule
\textbf{1} & \textbf{0}    & \textbf{No}  & \textbf{7722} & \textbf{905} & \textbf{8627} &\textbf{220} & \textbf{2416} & \textbf{12} & \textbf{2648} \\
2 & 0.10 & No  & 7822 & 910 & 8733 & 217 & 2429 & 12 & 2658 \\
3 & 0.20 & No  & 7797 & 947 & 8743 & 219 & 2426 & 12 & 2657 \\
4 & 0    & Yes & 7780 & 927 & 8707 & 219 & 2429 & 10 & 2658 \\
5 & 0.10 & Yes & 7802 & 927 & 8730 & 219 & 2431 & 10 & 2661 \\
6 & 0.20 & Yes & 7846 & 952 & 8798 & 219 & 2427 & 11 & 2657 \\
\bottomrule
\end{tabular}
\end{table}

\section{Additional Generalisation Results}
\label{app:4}

Table~\ref{tab:app_generalisation} provides additional results for the generalisation experiments in Section~\ref{sec:4-6}. For learned approaches, results are averaged over five runs per dataset, while FIFO+RB is evaluated deterministically once per dataset. S0--S3 and S7--S8 are further averaged equally over the three held-out test datasets, whereas S4--S6 use their corresponding demand-variation datasets. Each reported column is rounded independently after aggregation.

Across the evaluated conditions, PPO-PL+TD3 consistently yields lower vehicle-service delay than FIFO+RB and generally outperforms SFGH+TD3. Relative to FIFO+RB, the reduction in $C^{svc}$ ranges from approximately 3.2\% under S0--S3 to 6.0\% under S8. Compared with SFGH+TD3, PPO-PL+TD3 achieves lower $C^{svc}$ in all scenarios except S4, where the difference is marginal. For energy dispatch, PPO-PL+TD3 consistently yields lower total energy cost than FIFO+RB. Relative to SFGH+TD3, PPO-PL+TD3 achieves lower $C^{to}$ in five of the eight unseen scenarios, with only marginal differences in the remaining cases. Overall, these results show that the proposed dual-agent PPO-PL+TD3 framework provides stronger generalisation than the fully heuristic baseline (FIFO+RB) and the single-agent-learning framework (SFGH+TD3), delivering lower vehicle-service delay and better overall energy-dispatch performance across the unseen conditions.
\begin{table*}[t]
\caption{Generalisation performance under S0--S8.}
\label{tab:app_generalisation}
\centering
\small
\setlength{\tabcolsep}{2pt}
\renewcommand{\arraystretch}{1.05}
\textbf{(a) Vehicle service (h)}\par\vspace{2pt}
\begin{tabular*}{\textwidth}{@{\extracolsep{\fill}}ll*{3}{rrr}@{}}
\toprule
 & & \multicolumn{3}{c}{\textbf{PPO-PL+TD3}} & \multicolumn{3}{c}{\textbf{FIFO+RB}} & \multicolumn{3}{c}{\textbf{SFGH+TD3}} \\
\cmidrule(lr){3-5} \cmidrule(lr){6-8} \cmidrule(lr){9-11}
\textbf{Scenario} & \textbf{Setting} & $C^{wt}$ & $C^{ot}$ & $C^{svc}$ & $C^{wt}$ & $C^{ot}$ & $C^{svc}$ & $C^{wt}$ & $C^{ot}$ & $C^{svc}$ \\
\midrule
S0 & Nominal & \textbf{7722} & \textbf{905} & \textbf{8627} & 7928 & 981 & 8910 & 7749 & 955 & 8704 \\
S1 & PV: 0\% & \textbf{7722} & \textbf{905} & \textbf{8627} & 7928 & 981 & 8910 & 7749 & 955 & 8704 \\
S2 & PV: 100\% & \textbf{7722} & \textbf{905} & \textbf{8627} & 7928 & 981 & 8910 & 7749 & 955 & 8704 \\
S3 & PV: 150\% & \textbf{7722} & \textbf{905} & \textbf{8627} & 7928 & 981 & 8910 & 7749 & 955 & 8704 \\
S4 & Arrivals $\downarrow$ & 626 & 2 & 628 & 653 & 2 & 654 & \textbf{625} & \textbf{2} & \textbf{626} \\
S5 & Demand $\downarrow$ & \textbf{2211} & \textbf{29} & \textbf{2240} & 2303 & 32 & 2334 & 2217 & 30 & 2247 \\
S6 & Both $\downarrow$ & \textbf{239} & \textbf{0} & \textbf{239} & 252 & 0 & 252 & \textbf{239} & \textbf{0} & \textbf{239} \\
S7 & Infra.: 6/4 & \textbf{1778} & \textbf{19} & \textbf{1796} & 1851 & 20 & 1871 & 1790 & 19 & 1809 \\
S8 & Infra.: 4/2 & \textbf{116647} & \textbf{92517} & \textbf{209163} & 123328 & 99086 & 222415 & 122918 & 98807 & 221725 \\
\bottomrule
\end{tabular*}
\par\vspace{6pt}
\textbf{(b) Energy dispatch (k\pounds)}\par\vspace{2pt}
\begin{tabular*}{\textwidth}{@{\extracolsep{\fill}}ll*{3}{rrrr}@{}}
\toprule
 & & \multicolumn{4}{c}{\textbf{PPO-PL+TD3}} & \multicolumn{4}{c}{\textbf{FIFO+RB}} & \multicolumn{4}{c}{\textbf{SFGH+TD3}} \\
\cmidrule(lr){3-6} \cmidrule(lr){7-10} \cmidrule(lr){11-14}
\textbf{Scenario} & \textbf{Setting} & $C^{gd}$ & $C^{pl}$ & $C^{pv}$ & $C^{to}$ & $C^{gd}$ & $C^{pl}$ & $C^{pv}$ & $C^{to}$ & $C^{gd}$ & $C^{pl}$ & $C^{pv}$ & $C^{to}$ \\
\midrule
S0 & Nominal & \textbf{220} & \textbf{2416} & \textbf{12} & \textbf{2648} & 223 & 2447 & 16 & 2686 & 225 & 2416 & 10 & 2651 \\
S1 & PV: 0\% & \textbf{296} & \textbf{2443} & \textbf{0} & \textbf{2739} & 304 & 2449 & 0 & 2753 & 295 & 2458 & 0 & 2753 \\
S2 & PV: 100\% & \textbf{195} & \textbf{2418} & \textbf{37} & \textbf{2649} & 197 & 2443 & 38 & 2678 & 192 & 2433 & 33 & 2659 \\
S3 & PV: 150\% & \textbf{177} & \textbf{2411} & \textbf{89} & \textbf{2677} & 180 & 2432 & 72 & 2684 & \textbf{175} & \textbf{2422} & \textbf{80} & \textbf{2677} \\
S4 & Arrivals $\downarrow$ & 131 & 1428 & 23 & 1582 & 129 & 1444 & 26 & 1599 & \textbf{128} & \textbf{1425} & \textbf{23} & \textbf{1576} \\
S5 & Demand $\downarrow$ & 179 & 1802 & 16 & 1996 & 179 & 1818 & 18 & 2015 & \textbf{176} & \textbf{1798} & \textbf{20} & \textbf{1993} \\
S6 & Both $\downarrow$ & \textbf{106} & \textbf{1052} & \textbf{29} & \textbf{1187} & 102 & 1064 & 29 & 1195 & 103 & 1059 & 27 & 1188 \\
S7 & Infra.: 6/4 & \textbf{223} & \textbf{2428} & \textbf{12} & \textbf{2663} & 225 & 2446 & 16 & 2687 & 220 & 2437 & 18 & 2675 \\
S8 & Infra.: 4/2 & \textbf{216} & \textbf{2378} & \textbf{11} & \textbf{2605} & 216 & 2402 & 15 & 2633 & 211 & 2374 & 18 & 2603 \\
\bottomrule
\end{tabular*}
\end{table*}

\end{document}